\documentclass[iop]{emulateapj}
\usepackage{subfigure}
\usepackage{natbib}
\usepackage{url}
\usepackage{appendix}

\begin{document}

\shorttitle{The Evolution of Structure in IRDCs}
\shortauthors{Battersby et al.}

\def\Msun{\hbox{M$_{\odot}$}}
\def\Lsun{\hbox{L$_{\odot}$}}
\def\kms{km~s$^{\rm -1}$}	
\def\hcop{HCO$^{+}$}
\def\n2hp{N$_{2}$H$^{+}$}
\def\micron{$\mu$m}
\def\13CO{$^{13}$CO}
\def\etamb{$\eta_{\rm mb}$}	
\def\Inu{I$_{\nu}$}
\def\kapnu{$\kappa _{\nu}$}
\def\ffore{f$_{\rm{fore}}$}
\def\tastar{T$_{A}^{*}$}
\def\nh3{NH$_{3}$}
\def\deg{$^{o}$}
\def\abnh3{$\chi$$_{NH_{3}}$}
\def\H2{H$_{2}$}
\def\arcsec{$^{\prime\prime}$}
\def\arcmin{$^{\prime}$}
\newcommand{\oneone}{(1,1)}
\newcommand{\twotwo}{(2,2)}
\newcommand{\fourfour}{(4,4)}
\input epsf

\title{The Onset of Massive Star Formation:  The Evolution of Temperature and Density Structure in an Infrared Dark Cloud}

\author{Cara Battersby\altaffilmark{1,2},
Adam Ginsburg\altaffilmark{3,2},
John Bally\altaffilmark{2},
Steve Longmore\altaffilmark{4},
Miranda Dunham\altaffilmark{5},
Jeremy Darling\altaffilmark{2}
}

\altaffiltext{1}{Harvard-Smithsonian Center for Astrophysics, 60 Garden Street, Cambridge, MA 02138, USA}
\altaffiltext{2}{Center for Astrophysics and Space Astronomy, University of Colorado, UCB 389, Boulder, CO 80309}
\altaffiltext{3}{European Southern Observatory, Karl-Schwarzschild-Strasse 2, D-85748 Garching bei M¬unchen, Germany}
\altaffiltext{4}{Astrophysics Research Institute, Liverpool John Moores University, Twelve Quays House, Egerton Wharf, Birkenhead CH41 1LD, UK}
\altaffiltext{5}{Department of Astronomy, Yale University, New Haven, CT 06520}

\begin{abstract}
We present new \nh3 (1,1), (2,2), and (4,4) observations from the Karl G. Jansky Very Large Array (VLA) compiled with work in the literature to explore the range of conditions observed in young, massive star-forming regions.  To sample the effects of evolution independent from those of distance/resolution, abundance, and large-scale environment, we compare clumps in different evolutionary stages within a single Infrared Dark Cloud (IRDC), G32.02+0.06.  We find that the early stages of clustered star formation are characterized by dense, parsec-scale filamentary structures interspersed with complexes of dense cores ($<$0.1 pc cores clustered in complexes separated by $\sim$1 pc) with masses from about 10 to 100 \Msun. The most quiescent core is the most extended while the star forming cores are denser and more compact, showing very similar column density structure before and shortly after the onset of massive star formation, with peak surface densities $\Sigma$ $\gtrsim$ 1 g cm$^{-2}$. Quiescent cores and filaments show smoothly varying temperatures from 10-20 K, rising to over 40 K in star-forming cores.  We calculate virial parameters for 16 cores and find that the level of support provided by turbulence is generally insufficient to support them against gravitational collapse ($\langle$$\alpha_{\rm{vir}}$$\rangle$ $\sim$ 0.6). The star-forming filaments show smooth velocity fields, punctuated by discontinuities at the sites of active star formation. 
We discuss the Massive Molecular Filament (MMF; M $\sim$ 10$^{5}$ \Msun, length $>$ 60 pc) hosting the IRDC, hypothesizing that it may have been shaped by previous generations of massive stars.
\end{abstract}


\section{Introduction}
The light from massive stars (M $\ge$ 8 \Msun) and stellar clusters dominates our picture of the universe.  Through their ionizing radiation, powerful winds and outflows, and explosive deaths, they have molded their galactic environment and defined the life cycles of gas and stars around them.  However, much remains to be understood concerning the formation processes of stars in a clustered environment.  Are there required initial conditions for the onset of massive star formation in our Galaxy or can they form in a range of  environments?  

One first step to addressing these questions observationally is to identify locations where massive stars are forming.  It is thought that most, if not all, massive stars form in a clustered environment \citep{lad03, dew05}.  Infrared Dark Clouds (IRDCs) are likely candidates, as their cold ($<$ 20 K), high column density (N(H$_{2}$) $>$ 10$^{22}$ cm$^{-2}$), and high-mass (10$^{2}$ - 10$^{4}$ \Msun) nature are ideally suited for forming 
 massive stars and stellar clusters \citep{car00, rat06, bat10}, though as \citet{kau10} point out many are insufficiently massive to do so.  Galactic Plane Surveys \citep[e.g., Bolocam Galactic Plane Survey - BGPS, APEX Telescope Large Area Survey of the GALaxy - ATLASGAL, and the Herschel Infrared Galactic Plane Survey - Hi-GAL,][]{gin13, agu11, sch09, mol10} of dust continuum trace cold, optically thin dust throughout the Galaxy and corroborate the picture of massive stars and clusters forming in cold, high-density clouds.  However, ammonia observations of BGPS and ATLASGAL clumps have shown that they sample a range of evolutionary stages, from pre- to actively star-forming \citep[e.g.,][]{dun11b, wie12} and these groups have worked to isolate starless 
clumps\footnote[1]{In this work, we use the terminology wherein clouds are roughly 10$^{3}$ - 10$^{4}$ \Msun~with sizes of 2-15 pc, clumps are roughly 10$^{2}$-10$^{3}$ \Msun~with sizes of 0.3-3 pc, and cores are roughly 1-100 \Msun~with sizes of about 0.03-0.2 pc as in \citet{ber07}.  We use the term `core complexes' to refer to clumps that are resolved into their sub-structured cores.} \cite[e.g.,][]{tac12, wil11}.  

The evolutionary stages of massive star-forming clumps have been categorized observationally by different groups \citep[e.g.;][]{bat10, cha09, pur09}.  Using a physically motivated model for the progression of massive star formation, \citet{bat10} organized the observational stages into four categories, from infrared dark and quiescent, to diffuse and actively star-forming.  Stage 1 objects are cold, dense clouds, often observed as IRDCs (if the viewing angle, IR background, etc. are preferable to do so) and show no signs of active star formation.  Stage 2 clumps have early signs of star formation evidenced by activity in shocks and outflows, like CH$_{3}$OH masers and Extended Green Objects \citep[EGOs;][]{cyg08, cha09}, but remain IR-dark.  In Stage 3, clumps have formed a massive star and begin to light up in the IR (and may show signatures of an UCHII region) and may still be accreting.  In Stage 4, the young UCHII region has expanded and blown out much of the natal stellar cocoon and appears as a diffuse mid-IR source.  \citet{bat11} showed that temperatures increase as a function of these observational signatures which may indicate a positive trend with evolution (temperature increases with time).  These evolutionary sequences connect our expected behavior of massive star forming regions with observational signatures, but there are uncertainties in this connection, such as core/outflow inclination, clump mass, and environment, particularly with a small sample size, which are discussed in more detail in \S \ref{sec:ev}.  Therefore, these evolutionary stages are used as a guideline only for observed signatures of massive star formation.

Ammonia has been shown to be a robust tracer of physical properties (particularly temperature) in star-forming regions \citep[e.g.][]{ho83, man92, lon07, juv12, pil06, pil11} and an excellent way to address the physical conditions associated with the onset of massive star formation.  Due to the hierarchical clustered nature of star formation \citep[e.g.,][]{lad03, gou12}, an understanding of the conditions of massive star formation requires both large-scale and high-resolution analyses of their physical properties.  In \citet{bat11}, we measure the physical conditions (dust temperatures and column densities) of massive star forming regions on about 1 pc scales, and in this work we compare this global picture with high-resolution observations.  Various high-resolution observations toward young clustered star-forming regions \citep[e.g.,][]{bro11, bro09, fon12, li12, rat08} begin to reveal the nature of   
star-forming protoclusters, hot molecular cores, and chemical variation.  High-resolution ammonia observations \citep[e.g.,][]{wan08, zha11, liu12, dev11} have revealed a strong correlation between the emission of \nh3 and dust emission and given some insight into the properties of young massive star forming regions.  In this paper, we probe changes in the physical properties of massive star-forming regions as a function of evolution.  

In this work, we present high-resolution ($\sim$ 0.1 pc) 
VLA observations of 3 para inversion transitions of \nh3 toward an IRDC containing sub-regions in a range of evolutionary states.  We eliminate common observational biases by observing these phases within a single IRDC.  Therefore, the observed clumps share a common distance, cloud properties, and large-scale environment.  We perform radiative transfer modeling on the \nh3 lines to produce maps of the temperature, column density, velocity, and velocity dispersion.  We present these observations and compare with previous results to summarize our understanding of the structure of massive star forming regions and how they evolve.  In a companion paper \citep{bat14a}, we compare the gas properties (temperatures and column densities) with those derived from dust and derive an abundance of \nh3 in this IRDC.

In \S \ref{sec:obs} we present the observations, data reduction, and source selection and in \S \ref{sec:nh3model} discuss the radiative transfer model.  In \S \ref{sec:results} we present the temperature, column density, and velocity field maps derived for this IRDC and identify a young HII region.  We discuss fragmentation and evolutionary stages in \S \ref{sec:frag} and \S \ref{sec:ev}.  
  In \S \ref{sec:litcomp}, we compare the structure we observe with previous observations to explore the range of properties observed in massive star-forming regions.  We explore the large-scale properties of the IRDC and the Massive Molecular Filament in which it is embedded in \S \ref{sec:largescale}.
We summarize our conclusions in \S \ref{sec:conclusion}.

\begin{figure*}
\centering
\includegraphics[scale=0.8]{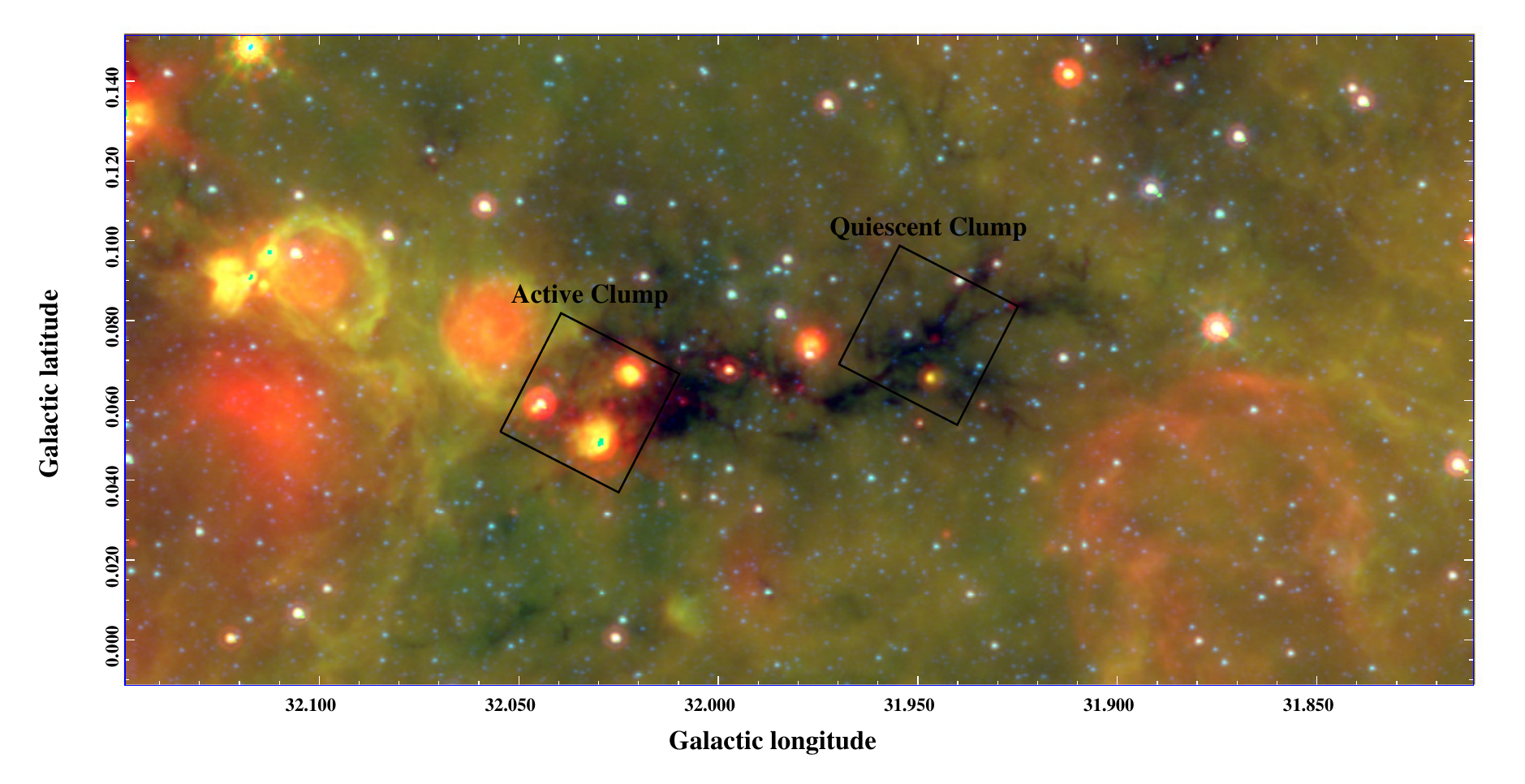}
\caption{The IRDC observed, G32.02$+$0.06, embedded within a massive molecular filament.  This IRDC shows different stages of massive star formation, from extended HII regions to infrared dark, cold gas.  The two black squares highlight the active and quiescent clumps depicted throughout the remainder of the paper.
The background is a Spitzer 3 color image, Red: MIPSGAL 24 \micron, Green and Blue: GLIMPSE 8 \micron~and 4.5 \micron.  }
\label{fig:irdc_medsm_glm3color}
\end{figure*}

\begin{figure*}
\centering
\subfigure{
\label{fig:cold_glm4_contours}
\includegraphics[scale=0.5]{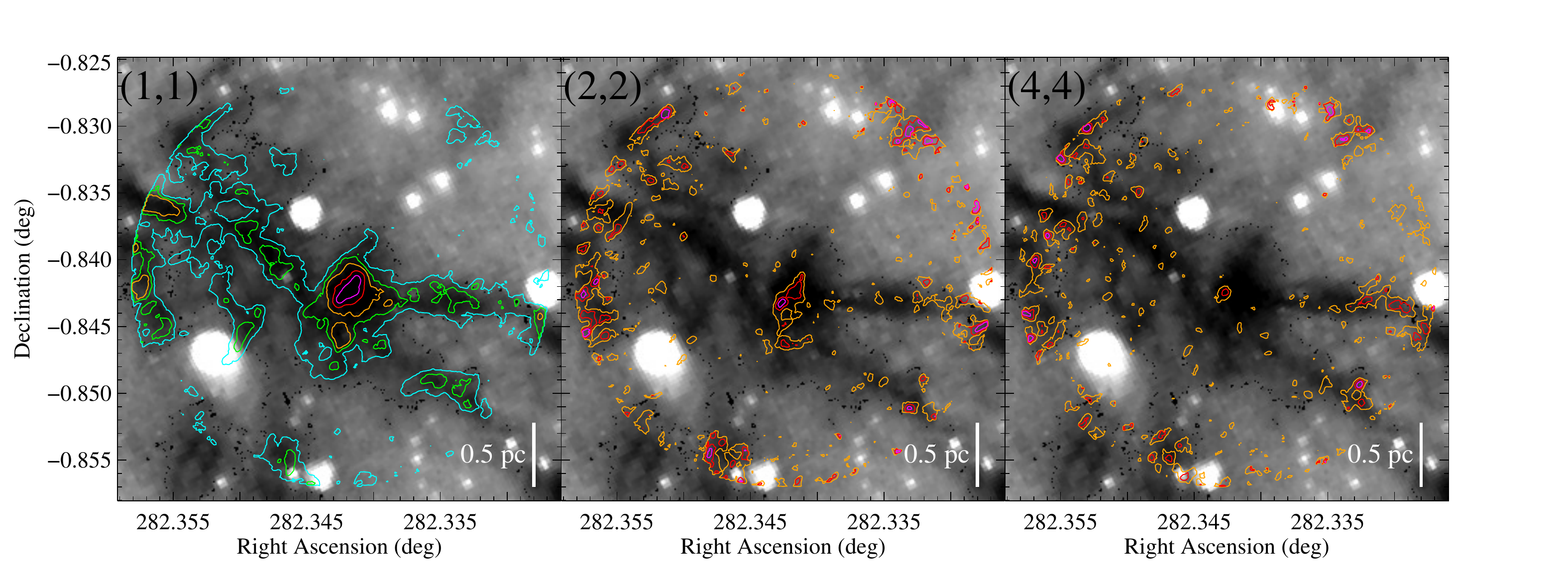}}\\
\vspace{-0.4in}
\subfigure{
\label{fig:hot_glm4_contours}
\includegraphics[scale=0.5]{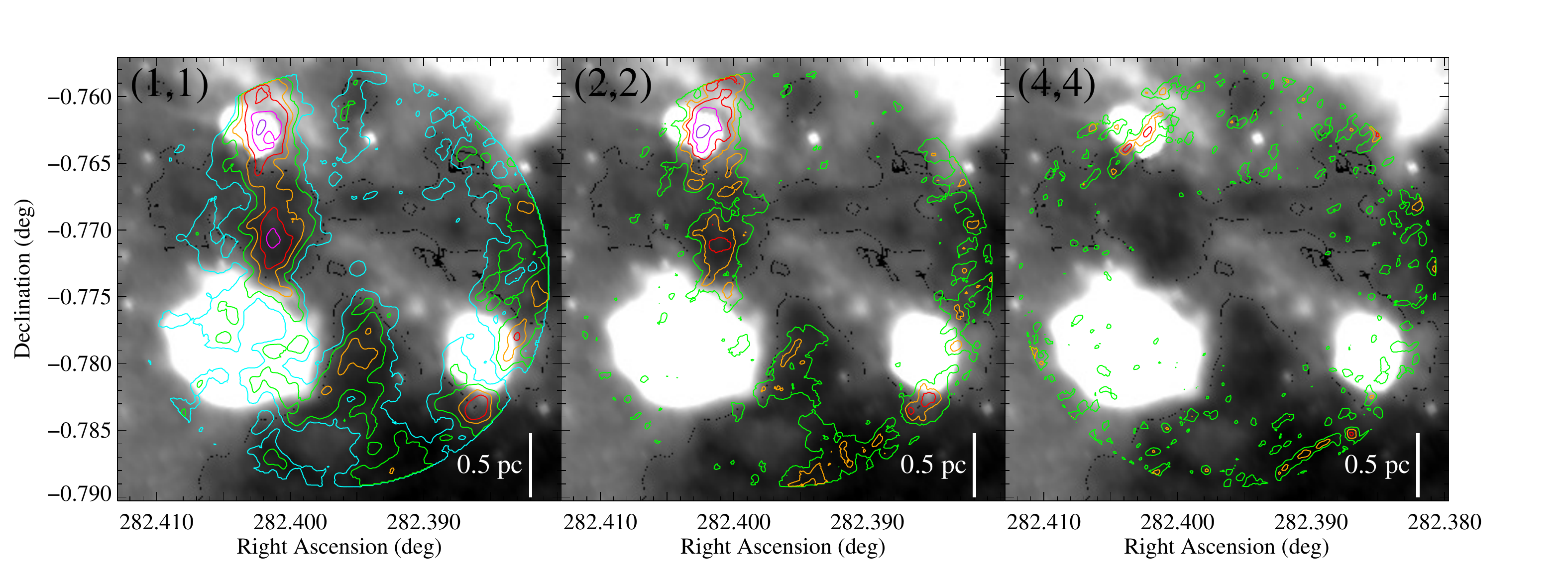}}\\
\caption 
{G32.02$+$0.06: GLIMPSE 8 \micron~grayscale image of the quiescent clump (\textit{top}) and the active clump (\textit{bottom}) with \nh3 contours.  Note the excellent morphological agreement between the 8 \micron~extinction and the \nh3 emission.  The quiescent clump shows no signs of massive star formation, while the active clump contains a range of star formation activity, from a young UCHII region that has blown out much of the dense gas and dust (the 8 \micron~bubble in the lower left of the image), to a cluster of cores showing 8 and 24 \micron~emission (in the top left) and an 8 \micron~dark complex in between.  The lowest contour plotted in each panel corresponds to about 1 $\sigma$.  \textit{In the quiescent clump in the top panel: Left: } \nh3 (1,1) integrated intensity contours [10,20,30,50,70 mJy/beam \kms].  \textit{Middle:} \nh3 (2,2) integrated intensity contours [30,50,70 mJy/beam \kms]. \textit{Right:} \nh3 (4,4) integrated intensity contours [30,50,70 mJy/beam \kms].  \textit{In the active clump in the bottom panel: Left:} \nh3 (1,1) integrated intensity contours [20,40,70,100,150,200 mJy/beam \kms].  \textit{Middle:} \nh3 (2,2) integrated intensity contours [30,50,70 mJy/beam \kms]. \textit{Right:} \nh3 (4,4) integrated intensity contours [40,70,100,150,200 mJy/beam \kms].
}
\label{fig:glm4_contours}
\end{figure*}

\section{Data}
\label{sec:obs}

\subsection{Source Selection}
\label{sec:selection}

We observed two clump locations within IRDC G32.02$+$0.06, an active clump 
([$\ell$, b] = [32.032\deg, $+$0.059\deg]) and a quiescent clump ([$\ell$, b] = [31.947\deg, $+$0.076\deg]) at a distance of $\sim$ 5.5 kpc (see next paragraph). 
The two clumps are embedded within a filamentary IRDC which is part of a much larger Massive Molecular Filament \citep[MMF, see \S \ref{sec:largescale} and][]{bat12a}, G32.02$+$0.06, see Figure \ref{fig:irdc_medsm_glm3color}.  The MMF is about 80 pc long \citep[as measured in \13CO with the Galactic Ring Survey,][]{jac06} in the Galactic mid-plane with a total mass of about 2 $\times$ 10$^{5}$ \Msun.  The velocity gradient across G32.02$+$0.06 is less than 4 \kms, and it appears to have been formed at the intersection of two UV-driven bubbles from previous generations of massive stars.  

The IRDC has V$_{LSR}$ = 95 \kms~\citep{dun11b}.  As IRDCs are seen in absorption against the bright mid-IR background, we assume the near kinematic distance for this cloud d $\sim$ 5.5 kpc \citep[assuming the rotation curve of][]{rei09}.  We note that some IRDCs can be located at the far distance \citep{ell13, bat11}, but that this IRDC is too dark to be consistent with the far kinematic distance.  At a distance of 5.5 kpc, the beam size of the observations is approximately 0.07 pc (beam dimensions given in \S \ref{sec:obsdet}), and the field of view for each clump is about 3 $\times$ 3 pc.  The active clump has a mass of $\sim 5000-10000$ \Msun~\citep[total mass calculated by summing the Hi-GAL column density map derived in companion paper][]{bat14a}, and displays signs of active star formation; a 6.7 GHz methanol maser \citep{pes05}, 8 and 24 \micron~emission indicative of a handful of Ultra-Compact HII Regions \citep[``Diffuse Red Clump" and 24 \micron~emission $\gtrsim$ 1 Jy][]{bat10}, and radio continuum emission \citep{whi05, hel06}.  The quiescent clump has a mass of about 3000 \Msun~and shows very little to no signs of active star formation (the dense gas may be associated with a faint 24 \micron~point source near the clump center), and is seen in absorption at 8 \micron.

\subsection{Observations}
\label{sec:obsdet}
We completed K-band observations of the (J,K) = (1,1), (2,2), and (4,4) inversion transitions of para-\nh3 over two tracks in the D configuration on 1 and 6 July, 2010 at the National Radio Astronomy Observatory\footnote[1]{The National Radio Astronomy Observatory is a facility of the National Science Foundation operated under cooperative agreement by Associated Universities, Inc.} Karl G. Jansky Very Large Array (VLA) under project AG830.  We observed two clump locations within IRDC G32.02$+$0.06, an active clump and a quiescent clump (with the presence/lack of star formation tracers as described in \S \ref{sec:selection}).  Pointing, bandpass and flux calibrations were performed on observations of J1331+3030 (3C286) at the beginning of the shift.  Pointing was performed twice more during each 3.5 hour observing shift on J1851+005.  Phase calibration was done on J1851+005, a 1.1 Jy K-band calibrator 1.5\deg~from the IRDC, every 10 minutes, between quiescent and active clump observations. 

The para-\nh3 (1,1) inversion transition (hereafter (1,1) transition) at 23694.4955 MHz was observed in the WIDAR OSRO2 mode with 1 subband/dual polarization mode with 8 MHz bandwidth and 256 channels for a resolution of 0.4 \kms.  The para-\nh3 (2,2) and (4,4) inversion transitions (hereafter (2,2) and (4,4) transitions) at 23722.599 MHz and 24139.352 MHz, respectively, were observed simultaneously in the WIDAR OSRO1 mode with 2 subbands/dual 
polarization mode with 4 MHz bandwidth and 64 channels each for a resolution of 0.8 \kms.  The synthesized beam diameters for the (1,1), (2,2), and (4,4) observations, respectively, were $\sim$3.6\arcsec~$\times$ 2.7\arcsec, $\sim$3.5\arcsec~$\times$ 3.0\arcsec, $\sim$3.5\arcsec~$\times$ 2.8\arcsec.  The beam position angle was about -40\deg.  The field of view for each clump was about 1.9\arcmin$\times$1.9\arcmin.  The final RMS noise in each image was approximately 6 mJy/beam (see \S \ref{sec:reduction} for details).

\subsection{Data Reduction and Masking}
\label{sec:reduction}
The reduced \nh3 (1,1), (2,2), and (4,4) images\footnote[2]{Fully reduced data cubes are available upon request.} are shown in Figure \ref{fig:glm4_contours}.  The VLA data were reduced using CASA, the Common Astronomy Software Applications package, version 3.3.0.  We used standard CASA routines for the flux, bandpass, gain, and phase calibration.  The quiescent clump showed no continuum emission, and the active clump had one strong continuum emitter (see \S \ref{sec:freefree} for details).  The UV continuum subtraction was done in the standard method (using the routine \texttt{uvcontsub}) for the quiescent clump and for the (1,1) transition of the active clump.  However, the (2,2) observations toward the active clump included a detection of the (2,2) hyperfine transition toward one core complex.  This prevented a traditional continuum subtraction, as the number of line-free channels was too few.  Therefore, we subtracted the continuum derived from the (1,1) active clump observation from the (2,2) and (4,4) cubes.

The continuum subtracted cubes were then cleaned using a standard setup with the CASA clean command.  We performed iterative masking while cleaning on all lines and both clumps except the (4,4) images whose signal was too  low, so we cleaned without a mask and increased the cleaning threshold from one to three sigma (sigma as determined from the dirty maps).  All of the images were primary beam corrected.  All the images are weighted using ``briggs" mode in CASA \citep{bri99}.  Spectra averaged over each core are shown in Figures \ref{fig:qspectra} and \ref{fig:aspectra}, while position velocity slices along the filamentary structures are shown in Figures \ref{fig:quiescent_pv} and \ref{fig:active_pv}.

The final RMS noise was calculated using the \texttt{immoments} routine in CASA on the line-free channels of the fully reduced, flux-corrected data cube.  We give the approximate RMS noise values at two points in the map; the map center and the 50\% power point of the primary beam.
The RMS noise for the (1,1) observations were approximately 5 mJy/beam per 0.4 \kms~channel in the center and 6 mJy/beam per 0.4 \kms~channel at the half-max of the primary beam. 
The (2,2) and (4,4) observations had the same RMS noise with values of approximately 6 mJy/beam per 0.8 \kms~channel in the center and 9 mJy/beam per 0.8 \kms~channel at the half-max of the primary beam.
A more sophisticated calculation of the noise parameters was done with the line fitting and temperature modelling, as discussed in \S \ref{sec:nh3model}.  Our continuum sensitivity in the (1,1) band is about 0.2 mJy.

Lastly, we applied a mask to each of the images in order to avoid image artifacts, low signal to noise, and kinematically unrelated regions.  The first version of the mask was created on each of the hot and cold clump (1,1) data cubes.  Using the integrated intensity, RMS, velocity, and velocity dispersion images  calculated from \texttt{immoments}, we  made a signal to noise cut of one on the (1,1) integrated intensity map.  The next cut was that the velocity dispersion must be less than 5 \kms~and that the velocity must be between 90 and 100 \kms, criteria which were easily met by the source signal.  Finally, the application of the primary beam corrections lead to extremely noisy regions at the edges of the maps, as well as enhancing image artifacts in these regions.  We exclude these regions by making a primary beam power point cutoff of 50\%.  These cutoffs were all performed on the (1,1) images, and constitute the final mask for the (1,1) images.  These masks were then applied to each of the (2,2) and (4,4) images with the added cutoff that the signal to noise, in the (2,2) and (4,4) maps respectively, must be greater than one.  These masks are then applied to the data and used throughout the remainder of this paper in the modeling, figures, and analyses.

\subsection{Archival Data}
To help understand the context and to determine some physical properties of this region, we utilize extant survey and literature data.  We use mid-IR data taken with Spitzer as part of the Galactic Legacy Infrared Mid-Plane Survey Extraordinaire \citep[GLIMPSE;][]{ben03} and 24 \micron~data taken as part of the MIPS Galactic Plane survey \citep[MIPSGAL][]{car09}.  We also use millimeter continuum data taken as part of the Bolocam Galactic Plane Survey\footnote[2]{http://irsa.ipac.caltech.edu/data/BOLOCAM\_GPS/} \citep[BGPS;][]{agu11, ros10, gin13} of the Galaxy at 1.1 mm .  The Boston University-Five College Radio Astronomy Galactic Ring Survey \citep[BU-FCRAO GRS;][]{jac06} provides spectral line data of \13CO J=1-0.  The catalogs from the HII Region Discovery Survey \citep[HRDS][]{and11, and12} were utilized as well as complementary radio continuum data (from the VLA) provided by the Multi-Array Galactic Plane Imaging Survey \citep[MAGPIS;][]{whi05, hel06} at 6 and 20 cm \citep[see also the newer CORNISH survey;][]{pur13, hoa12}.  
Additionally, we used the single dish \nh3 data taken on the Green Bank Telescope as part of a survey of BGPS sources by \citet{dun11b} in our original source selection.

We use dust continuum data taken as part of the Herschel Infrared Galactic Plane Survey \citep[Hi-GAL,][]{mol10}.  Data reduction was carried out using the \textit{Herschel} Interactive
Processing Environment \citep[HIPE,][]{ott10} with custom reduction scripts that deviated considerably from the standard processing for PACS \citep{pog10}, and to a lesser extent for SPIRE \citep{gri10}.  A more detailed description of the entire data reduction procedure can be found in \citet{tra11}.  A weighted post-processing on the maps \citep{pia13} has been applied to help with image artifact removal.

\section{Radiative Transfer NH$_{3}$ Modeling}
\label{sec:nh3model}
\begin{figure*}
  \centering
  \subfigure{
  \includegraphics[width=0.48\textwidth]{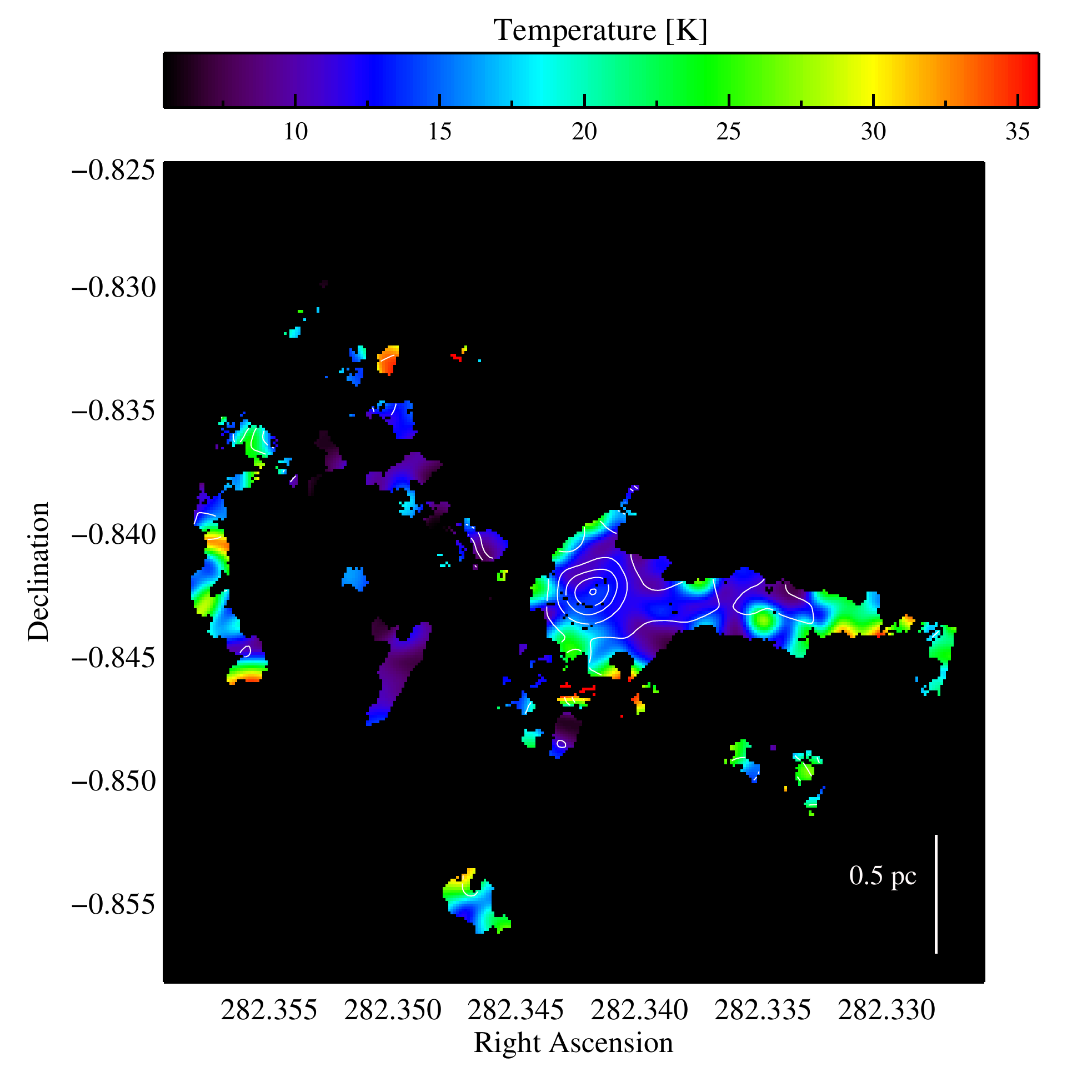}}
  \subfigure{
  \includegraphics[width=0.48\textwidth]{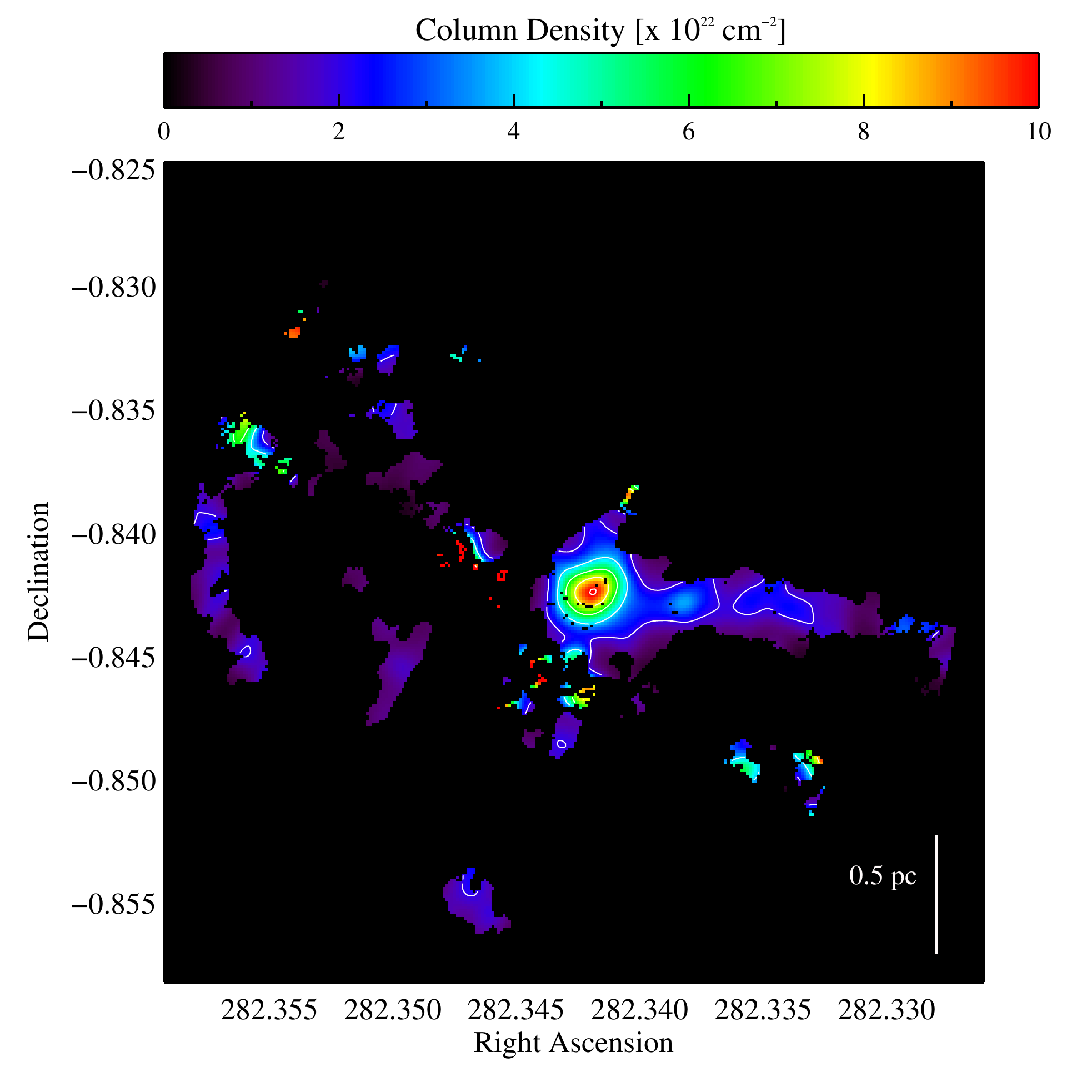}}\\
  \subfigure{
  \includegraphics[width=0.48\textwidth]{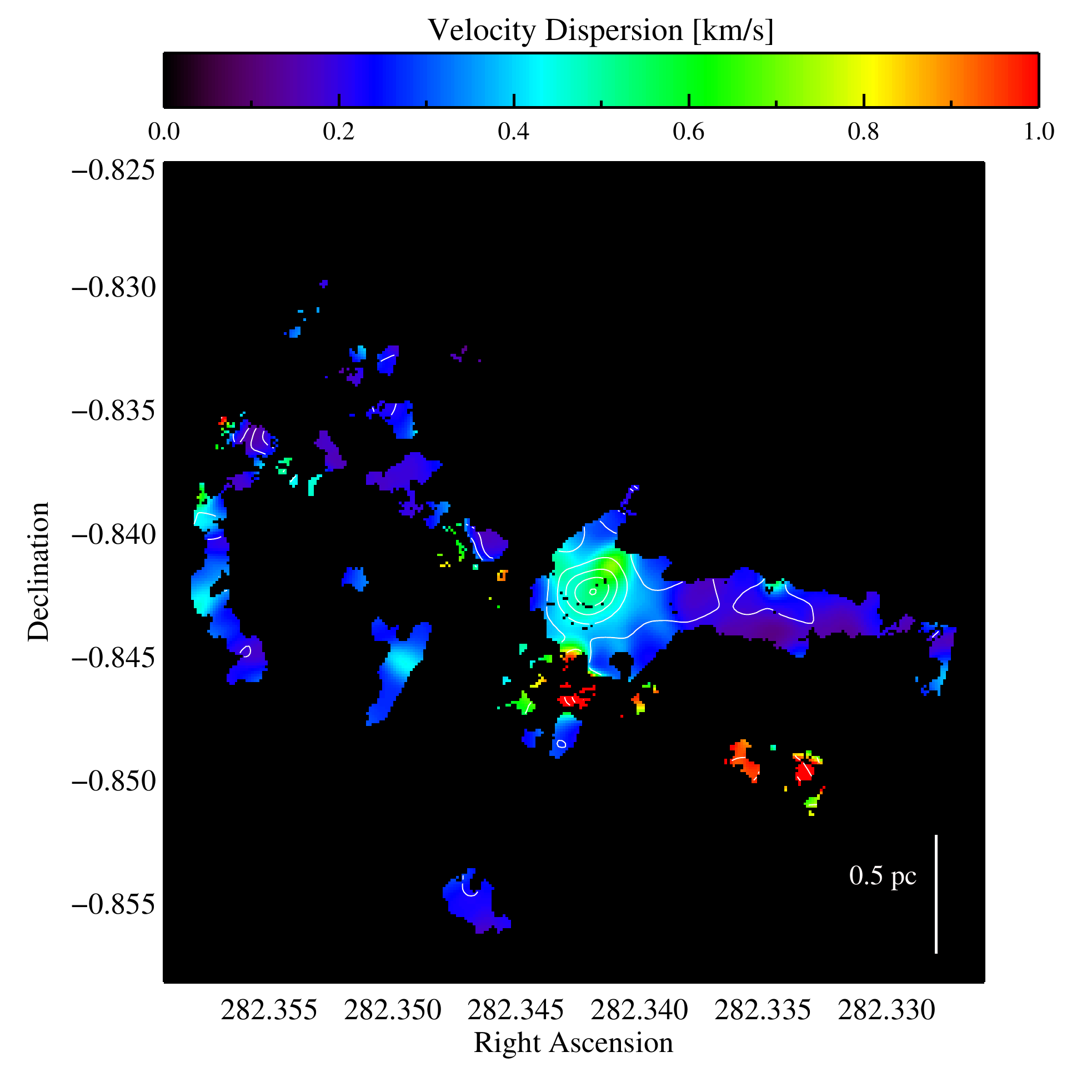}}
  \subfigure{
  \includegraphics[width=0.48\textwidth]{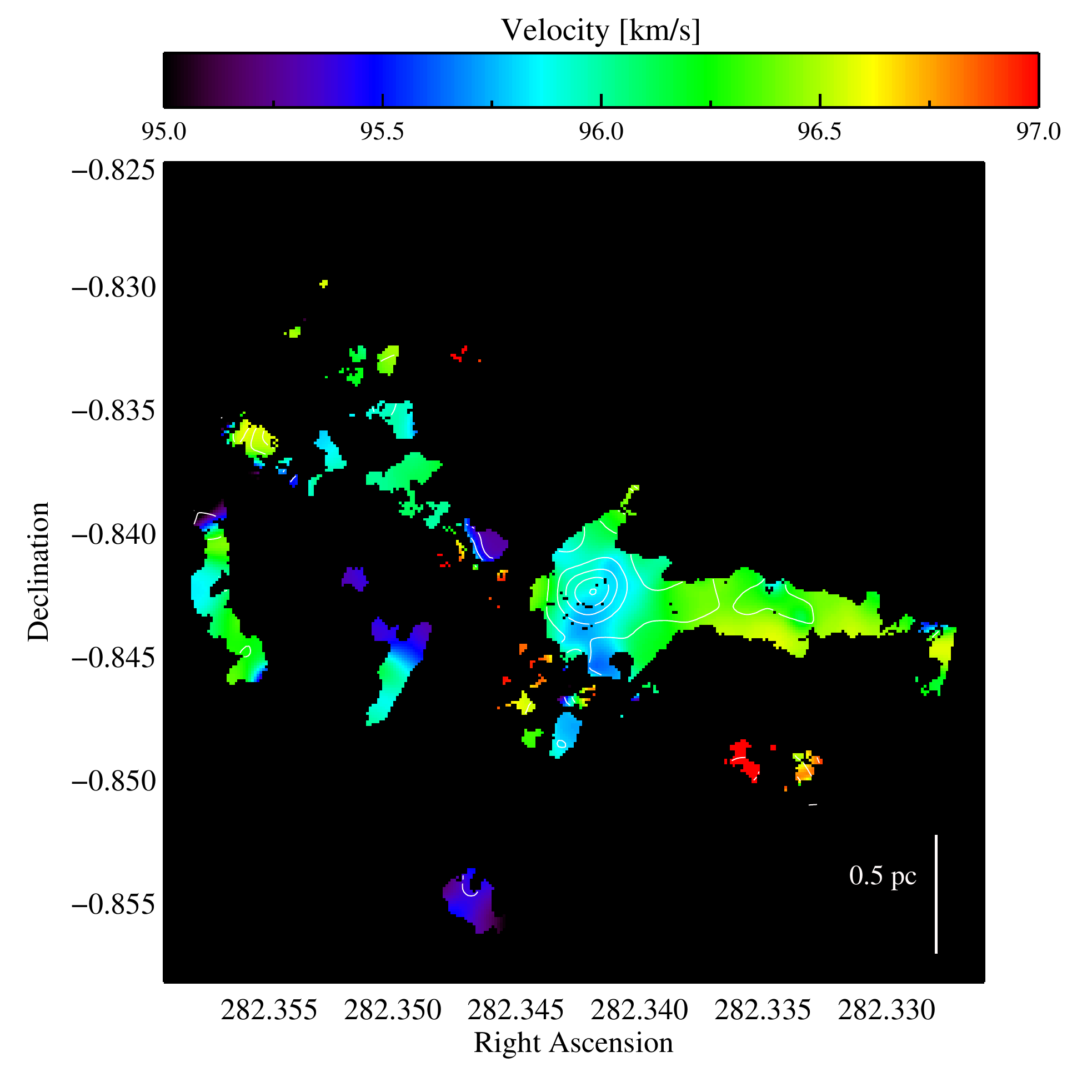}}\\
  \caption{``Low velocity'' component (integrated from 93 to 97.3 \kms) of the quiescent clump: Temperature, column density, velocity dispersion, and velocity (from top left to bottom right) maps as derived from the radiative transfer model fits (see \S \ref{sec:nh3model}).  The contours are N(\H2) = [2,4,6,8,10] $\times$ 10$^{22}$ cm$^{-2}$ assuming an abundance of 4.6 $\times$ 10$^{-8}$ from \citet{bat14a}.}
\label{fig:coldlow_fits}
\end{figure*}

\begin{figure*}
  \centering
  \subfigure{
  \includegraphics[width=0.48\textwidth]{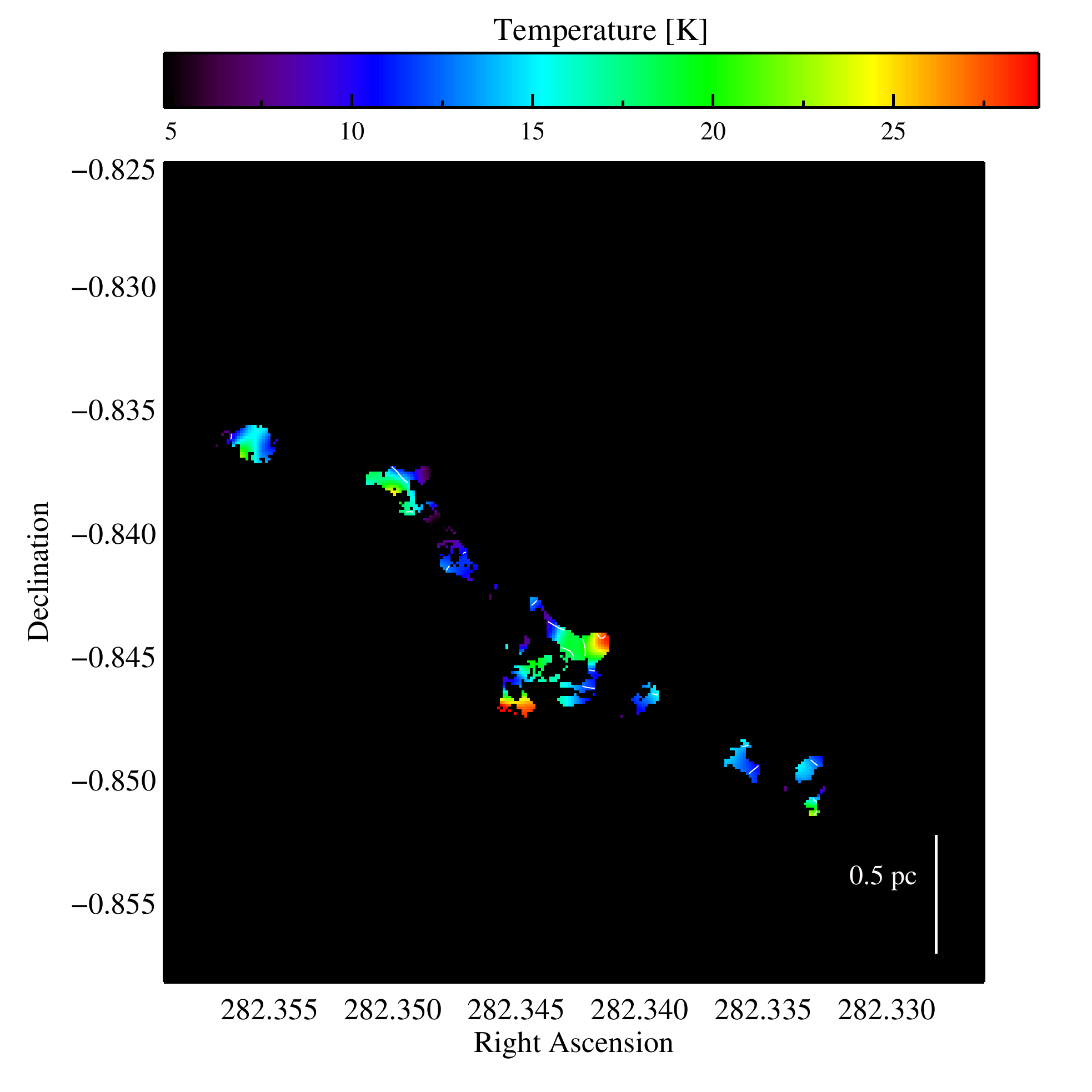}}
  \subfigure{
  \includegraphics[width=0.48\textwidth]{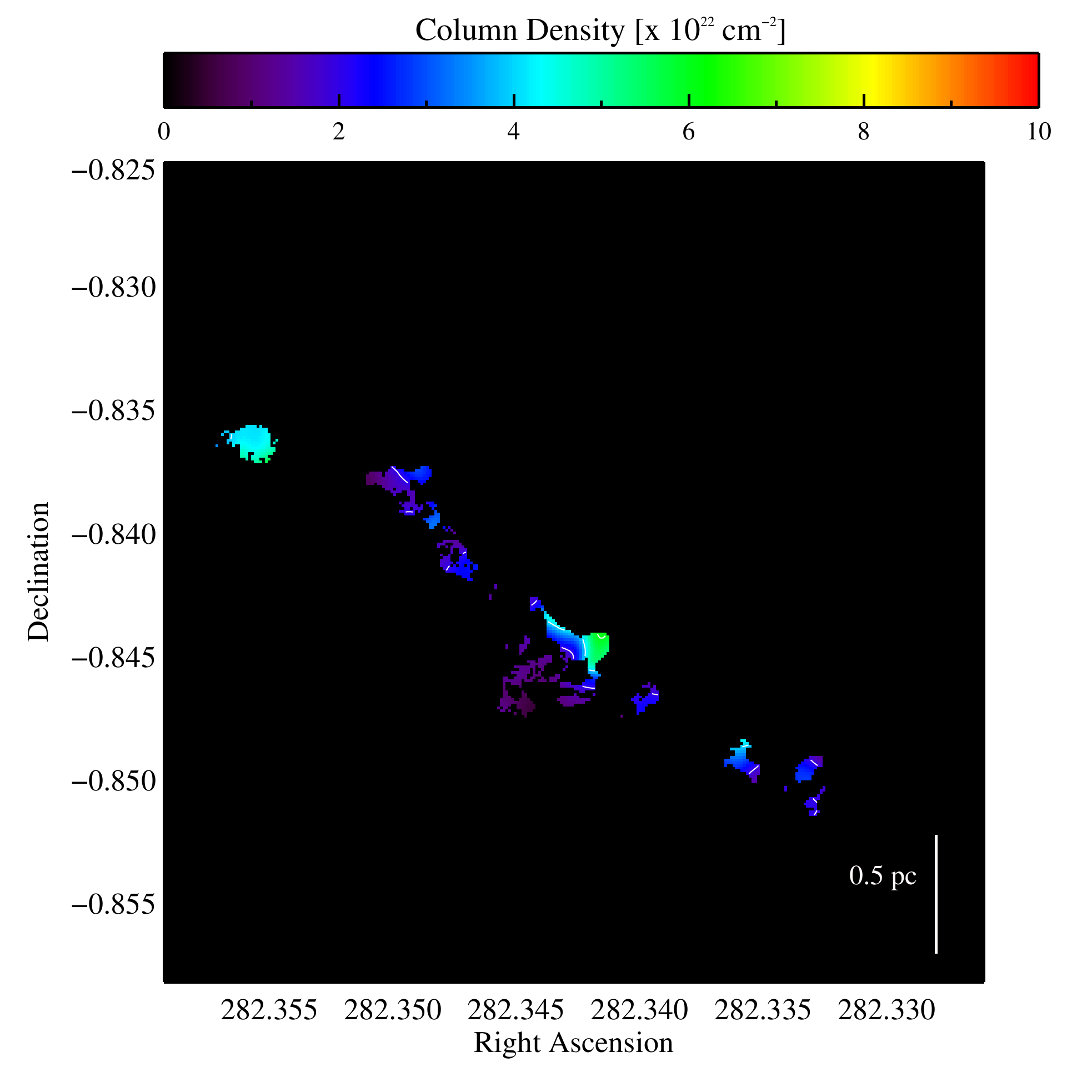}}\\
  \subfigure{
  \includegraphics[width=0.48\textwidth]{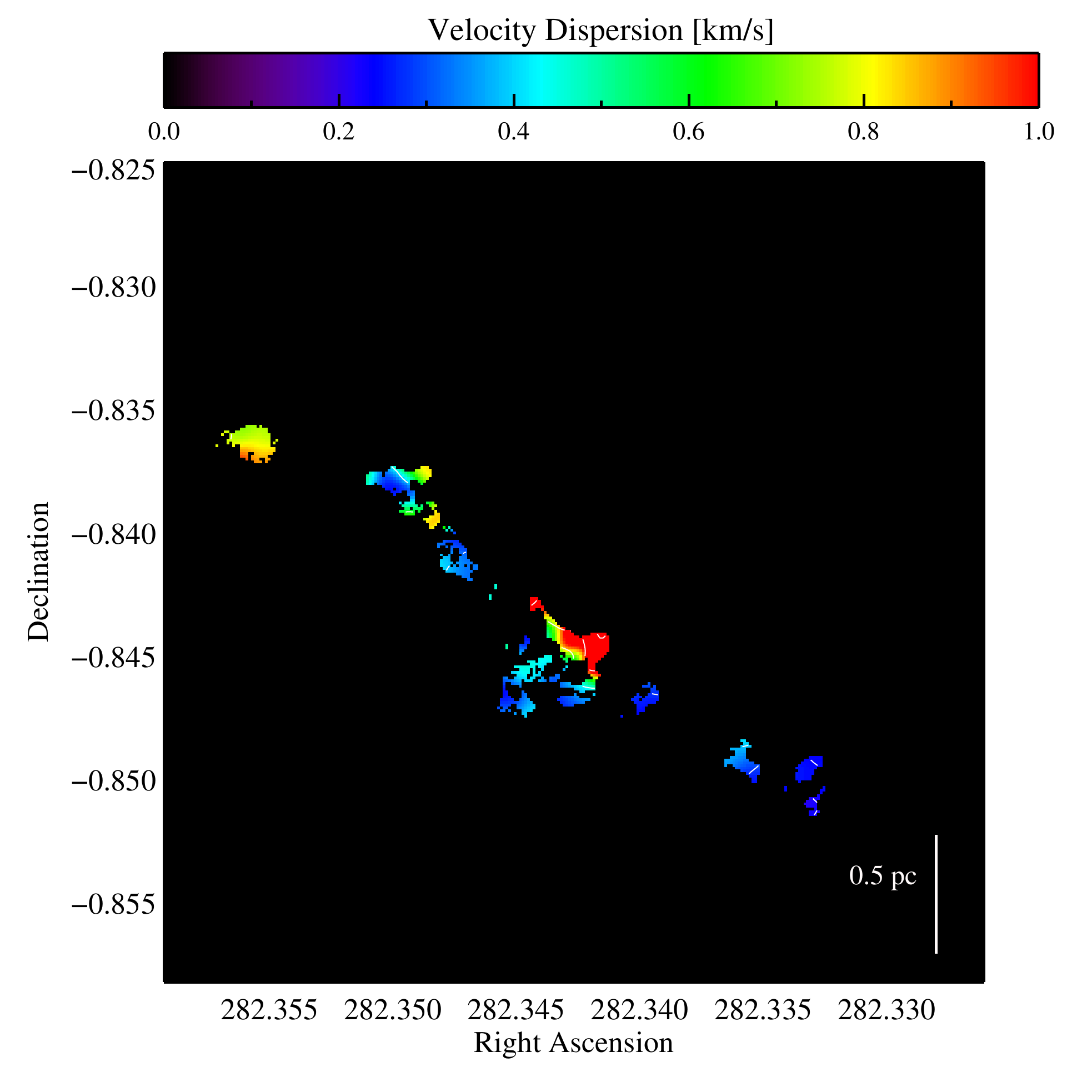}}
  \subfigure{
  \includegraphics[width=0.48\textwidth]{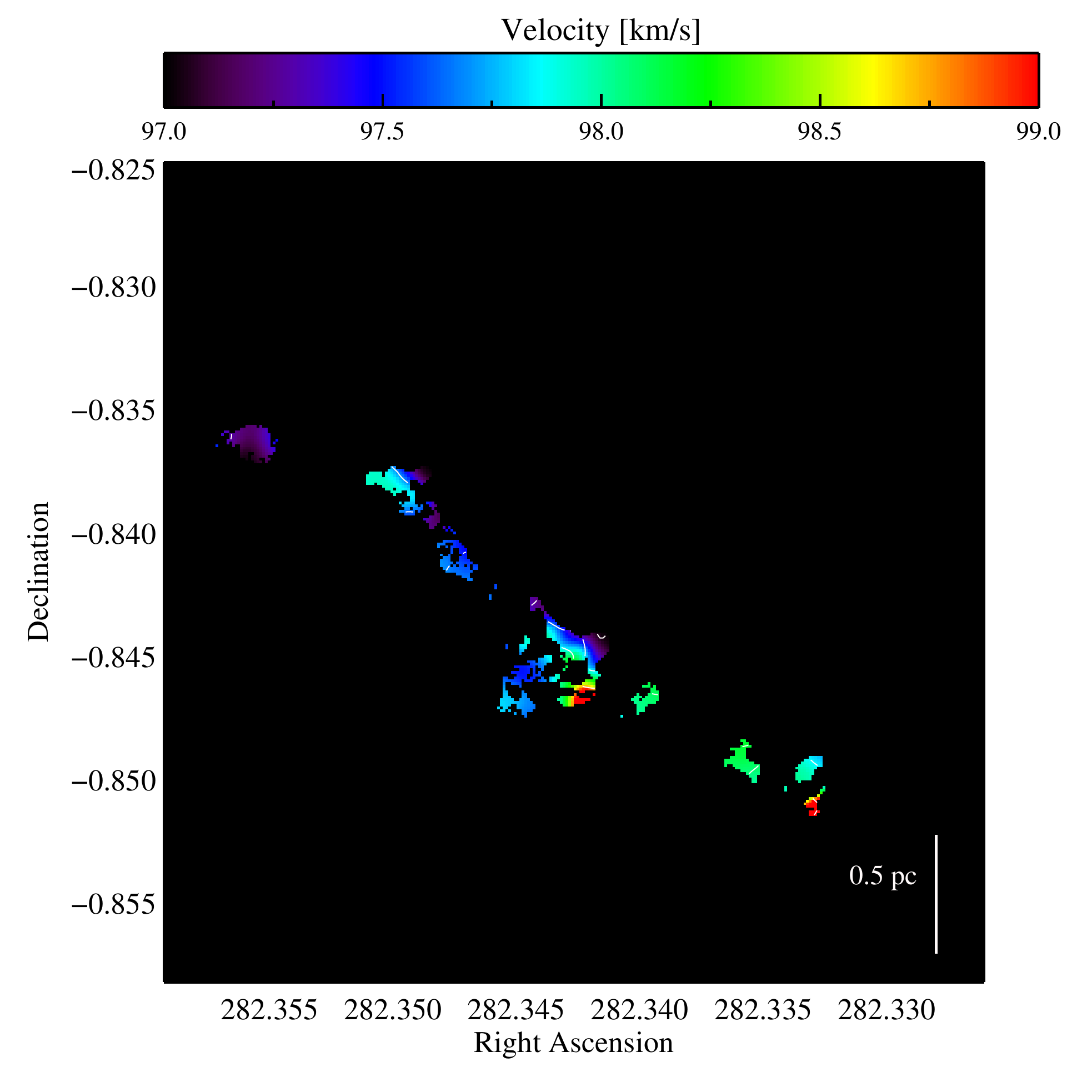}}\\
  \caption{``High velocity'' component (integrated from 97.3 to 101 \kms) of the quiescent clump: Temperature, column density, velocity dispersion, and velocity (from top left to bottom right) maps as derived from the radiative transfer model fits (see \S \ref{sec:nh3model}).  The contours are N(\H2) = [2,4,6,8,10] $\times$ 10$^{22}$ cm$^{-2}$ assuming an abundance of 4.6 $\times$ 10$^{-8}$ from \citet{bat14a}.}
  \label{fig:coldhigh_fits}
\end{figure*}

\begin{figure*}
  \centering
  \subfigure{
  \includegraphics[width=0.48\textwidth]{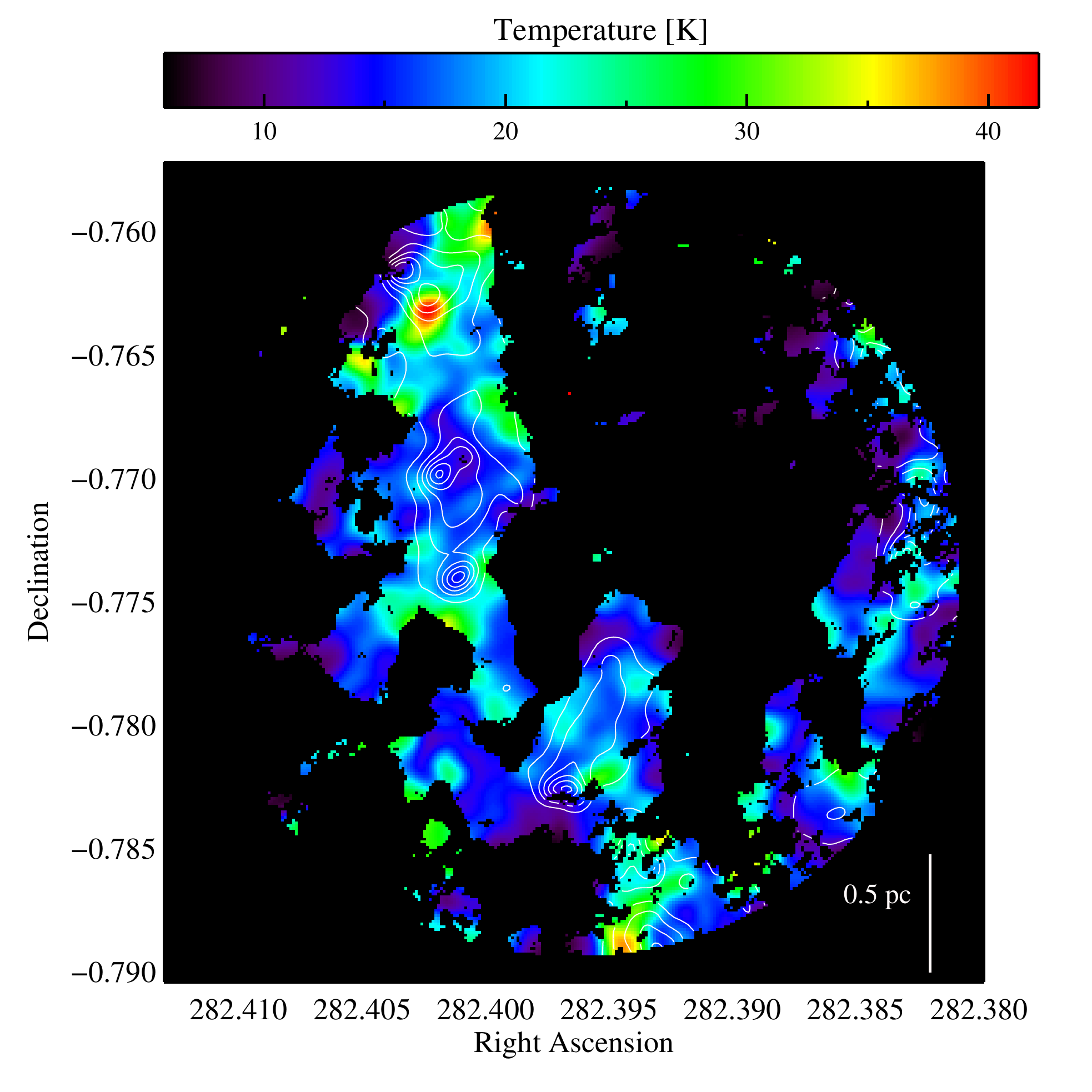}}
  \subfigure{
  \includegraphics[width=0.48\textwidth]{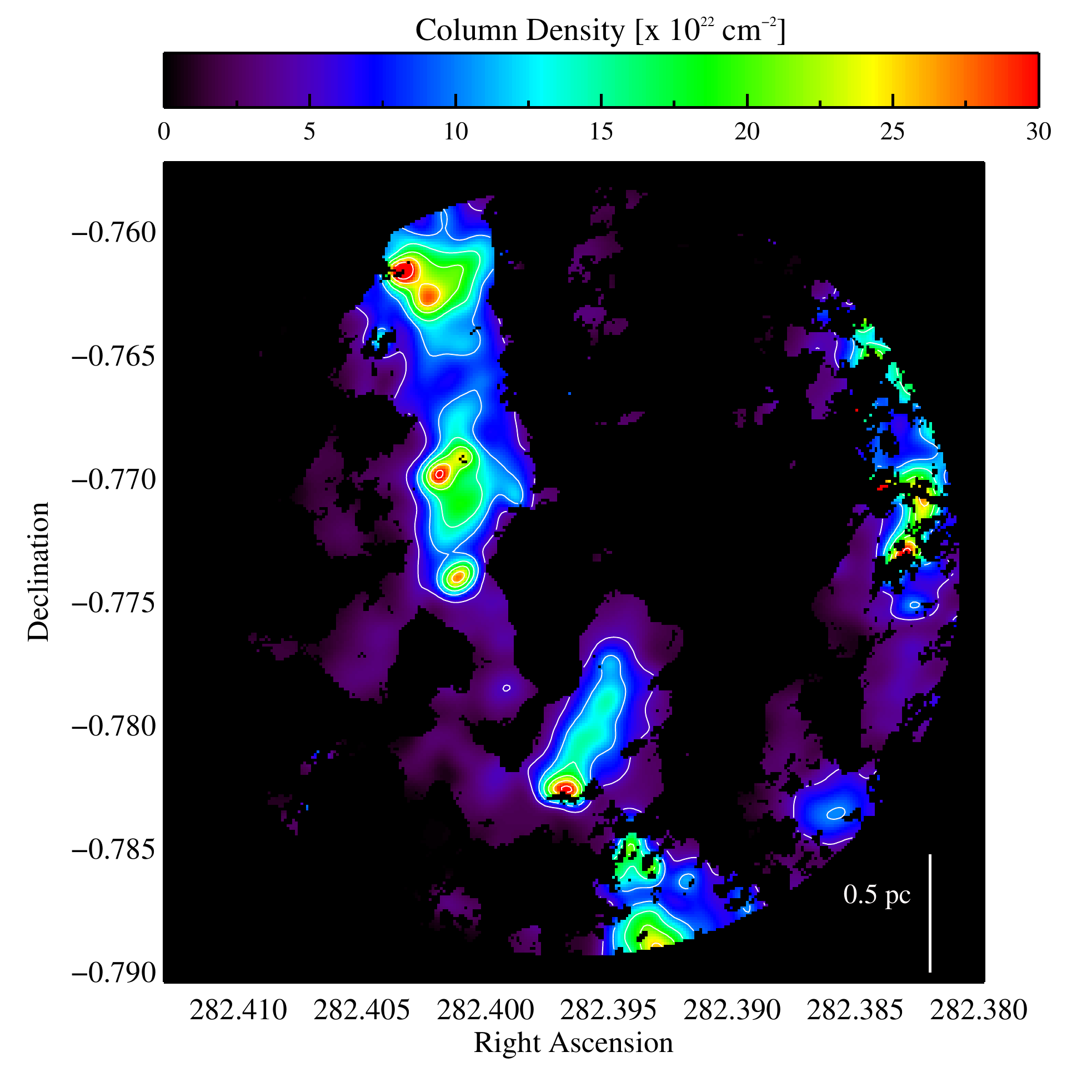}}\\
  \subfigure{
  \includegraphics[width=0.48\textwidth]{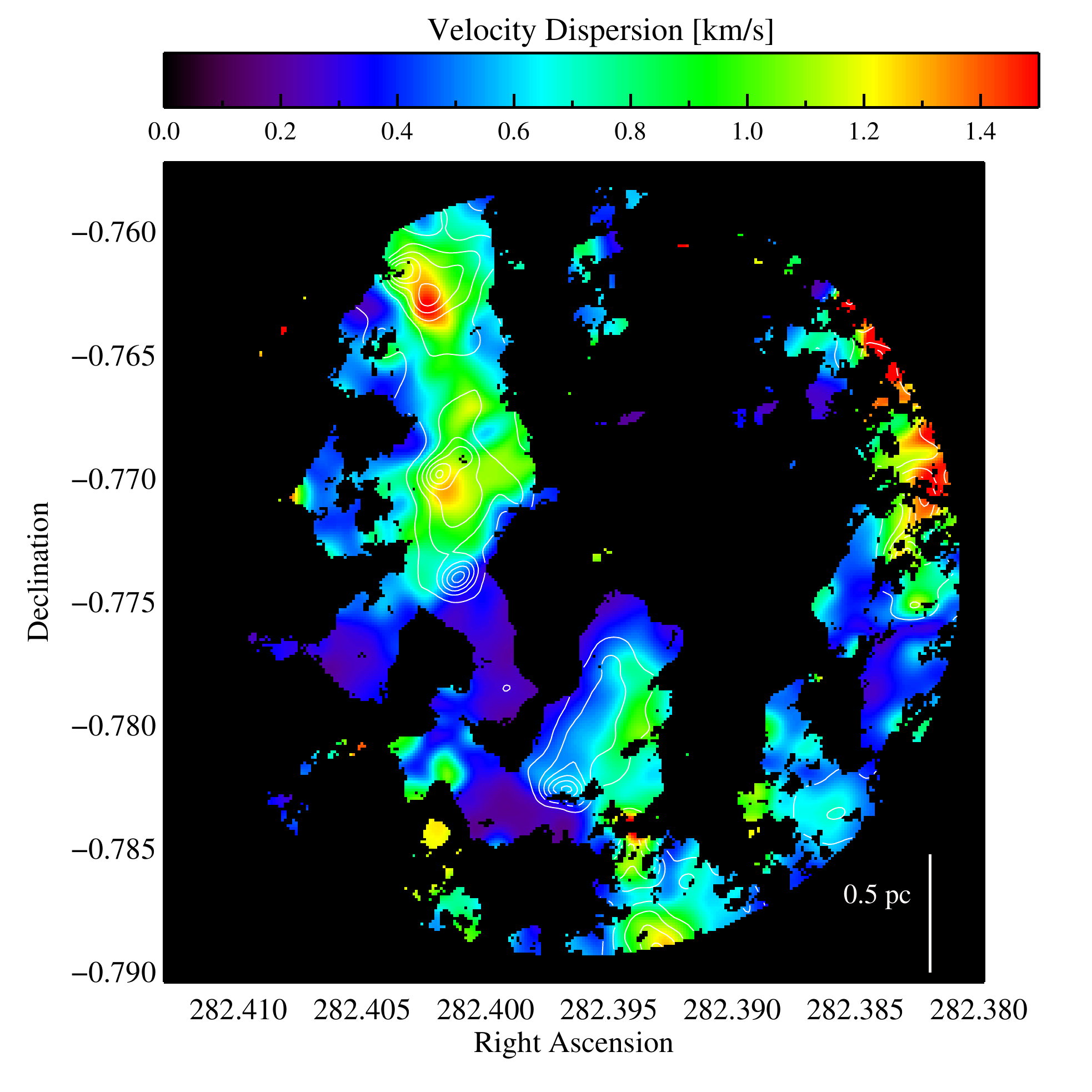}}
  \subfigure{
  \includegraphics[width=0.48\textwidth]{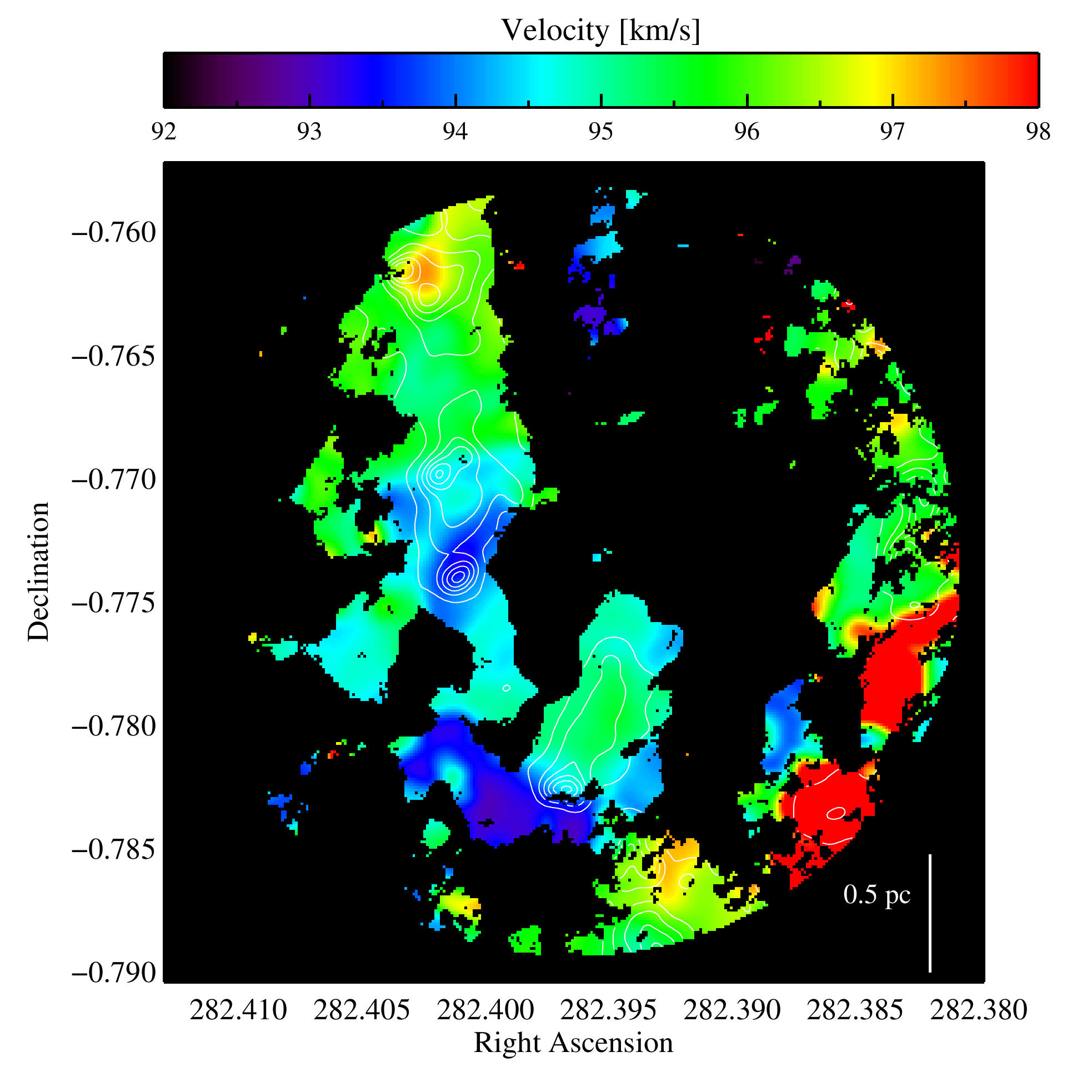}}\\
  \caption{Active clump: Temperature, column density, velocity dispersion, and velocity (from top left to bottom right) maps as derived from the radiative transfer model fits (see \S \ref{sec:nh3model}).  The contours are N(\H2) = [5,10,15,20,25,30] $\times$ 10$^{22}$ cm$^{-2}$ assuming an abundance of 4.6 $\times$ 10$^{-8}$ from \citet{bat14a}.}
  \label{fig:hot_fits}
\end{figure*}

The results of the radiative transfer \nh3 modeling are shown in Figures \ref{fig:coldlow_fits}, \ref{fig:coldhigh_fits}, and \ref{fig:hot_fits}. The NH$_3$ inversion transitions are a sensitive probe of temperatures in cold molecular gas \citep{ho83}.  The critical density of the (1,1) and (2,2) inversion transitions are similar at around $n_{crit}\approx10^{4.5}$ cm$^{-3}$, implying that \nh3 emission will only be detected in the densest part of GMCs most likely to form stars.  Throughout the text, we assume an abundance ($\frac{N(NH_{3})}{N(H_{2})}$), \abnh3  = 4.6 $\times$10$^{-8}$ derived using these data from \citet{bat14a}.

The ammonia lines were fitted with a Gaussian line profile to each hyperfine component simultaneously with frequency offsets fixed.  The fitting was performed in a python routine translated from Erik Rosolowsky's IDL fitting routines  \citep[Section 3 of ][]{ros08}, which forward models all of the spectral lines given input physical properties derived assuming a homogeneous slab of uniform  temperature, intrinsic velocity dispersion, and uniform excitation conditions for all the lines.
The model was used within the framework of the {\tt pyspeckit} spectral analysis code package \citep[][\url{http://pyspeckit.bitbucket.org}]{gin11}.  As in \citet{ros08}, the emission is assumed to arise from a homogeneous slab with uniform gas temperature, intrinsic velocity dispersion, and uniform excitation conditions for all hyperfine transitions of the \nh3 lines.  The model simultaneously fits the \oneone, \twotwo, and \fourfour~inversion transitions of para-\nh3 with 5 parameters: line excitation temperature $T_{ex}$, gas kinetic temperature $T_{kin}$, column density N(\nh3), velocity dispersion $\sigma$, and offset velocity $v$.  The models are fit using a Levenberg-Marquardt fitting code using {\tt scipy} (\url{http://scipy.org}, \url{https://github.com/newville/lmfit-py}).

Each pixel in the spectral data cubes was fitted independently.  The fits require an input guess as a starting point for the fitting algorithm.  The guess for each pixel was set to the nearest pixel with a valid fit to ensure continuity across the cube.  The first fit was performed by hand on a bright pixel, which was selected as the starting point for the loop.  The Levenberg-Marquardt algorithm computes a covariance matrix for the fitted parameters, but we report only the independent component of the errors on each parameter.  For optically thin lines, the column density and excitation temperature are degenerate, but they are fit independently so the data are treated consistently. Systematic offsets due to calibration errors could asymmetrically affect the temperature measurements, but because all three lines are observed simultaneously with the same calibration, such errors are expected to be small.

The fitted \oneone\ total optical depths ranged from $\tau\sim1$ (which is equivalent to $\tau\approx0.5$ for the central hyperfine line) to $\tau\approx80$ in some regions within the active clump (but see below) and $\tau\sim40$ in one small region within the quiescent clump.  For total \oneone\ optical depths $\tau>9$, the hyperfine lines have optical depths $\tau>1$.  Simulations show that the \oneone\ line can be used to accurately measure the observed properties (to within the reported error) up to $\tau \approx 20$, and the combination of \oneone\ and \twotwo\ are accurate up to the highest reported optical depths for kinetic temperatures above about 10 K because, for most combinations of optical depth and column density at which \nh3\ is detectable, either the \oneone\ hyperfine or the \twotwo\ main line are optically thin.  As noted in \citet{ros08}, $T_{ex}$ and column density are degenerate in the optically thin limit (i.e., where even the core \oneone\ line has $\tau \ll 1$), but this limit is essentially never reached at the sensitivity of our observations.

Typical statistical errors are in the range $\sigma$(T$_{K}$) $\sim$ 1-3 K for the kinetic temperature, and $\sigma\left(\log[N(NH_{3})]\right)\lesssim0.05$ (or about 10\%) 
for the column density of ammonia.  The statistical errors are representative if the gas within a beam is at a single temperature.  If there is gas at different temperatures within a single beam, the fitted temperature represents an average, but the errors do not reflect the underlying temperature distribution.  This caveat affects any line-of-sight temperature measurement, but should be less severe for \nh3\ than for dust temperatures because the \nh3\ temperatures typically select a single dense cloud, while dust temperatures average all clouds along a line of sight.

The fitting algorithm will return spurious or unreliable results when there are multiple velocity components along a line of sight.  We therefore re-fit parts of the data where multiple components were observed by restricting the velocities and fitting each component separately (See \S \ref{sec:modelfits}).  However, there are some `transition regions' where the two components are blended, and a single fit misrepresents both.  For additional information about caveats and uncertainties in \nh3\ line fitting, refer to \citet{ros08} Section 3.  

An additional mask (see \S \ref{sec:reduction}) is applied to the model fit images as applied throughout the remainder of the paper.  The only pixels that were excluded are obvious spurious bad model fits (e.g., a hot pixel) which were isolated spatially and therefore not physically reasonable.  These exclusions do not affect the peak map values reported in \S \ref{sec:results}.
For the quiescent clump, these excluded pixels had a temperature less than 5 K or greater than 50 K or an H$_{2}$ column density of less than 10$^{20}$ cm$^{-2}$ or greater than 2 $\times$ 10$^{23}$ cm$^{-2}$ or with a (1,1) integrated flux of less than 0.01 Jy/beam \kms.  For the active clump, these excluded pixels had a temperature less than 5 K or greater than 50 K or a column density of less than 10$^{20}$ cm$^{-2}$ or greater than 10$^{24}$ cm$^{-2}$ or with a (1,1) integrated flux of less than 0.02 Jy/beam \kms. 

%
\begin{figure*}
\centering
\subfigure{
\includegraphics[width=0.6\textwidth]{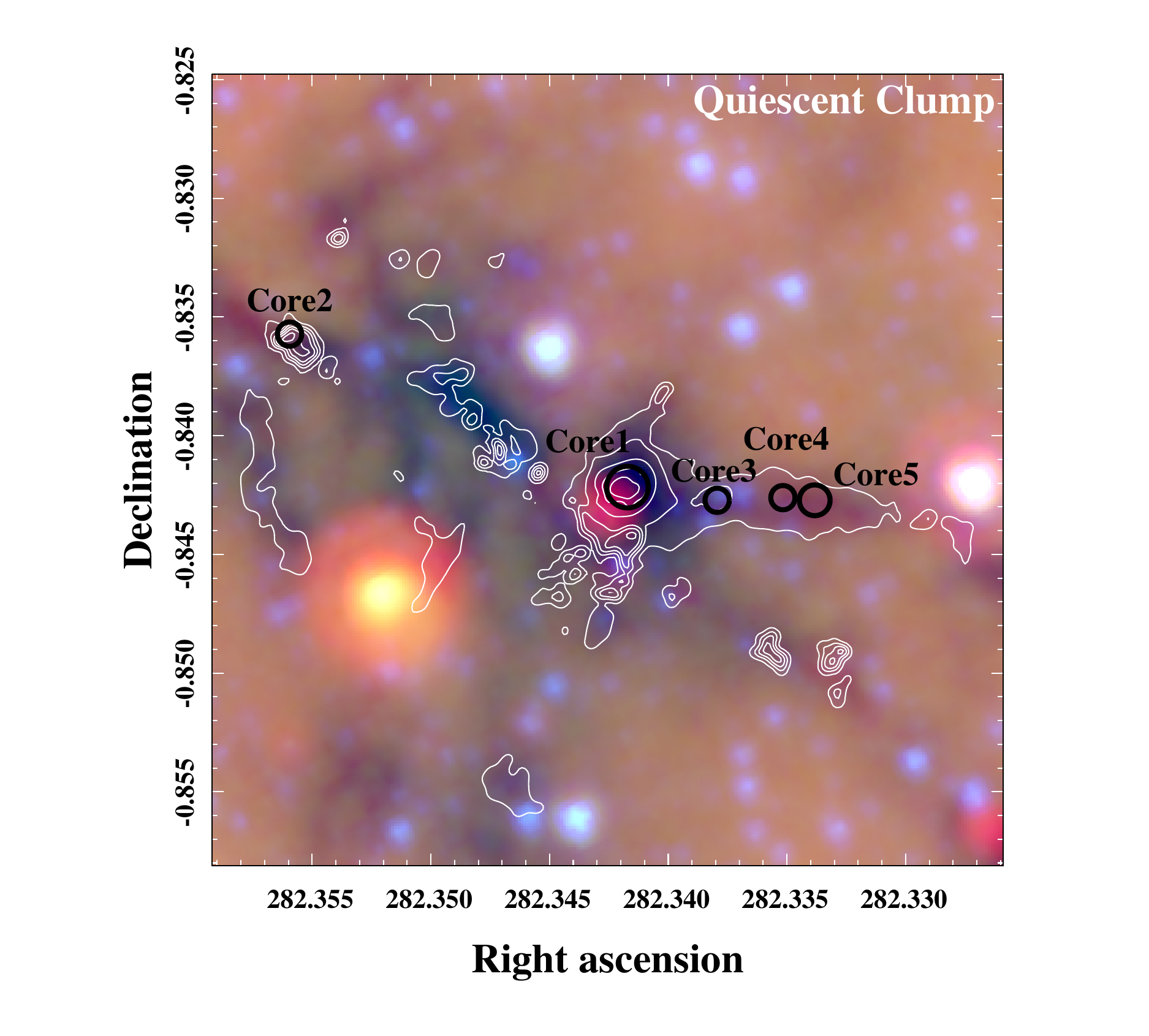}} \\
\subfigure{
\includegraphics[width=0.9\textwidth]{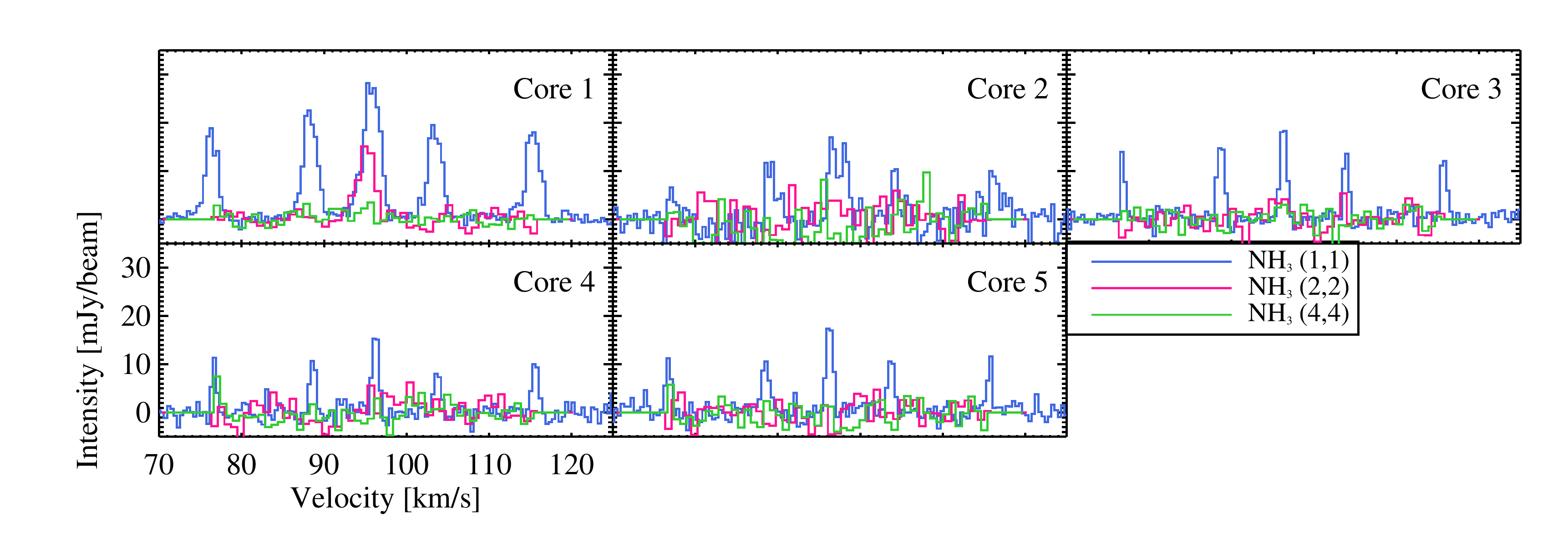}} \\
\caption{The \nh3 identified cores are labeled here for the quiescent clump.  \textit{Top:} Three color mid-IR image of each clump (MIPSGAL, GLIMPSE, Red: 24 \micron, Green: 8 \micron, Blue: 4.5 \micron) with column density contours derived from \nh3 (six linearly spaced contours from 8$\times$10$^{21}$ cm$^{-2}$ to 10$^{23}$ cm$^{-2}$).  The core sizes and locations are plotted as circles with the effective radii of the 2-D Gaussian fits.  \textit{Bottom:} \nh3 (1,1), (2,2), and (4,4) spectra averaged over each core. The brightness temperature conversion for these data is about 9 mJy/beam per 1 K. }
\label{fig:qspectra}
\end{figure*}

\begin{figure*}
\centering
\subfigure{
\includegraphics[width=0.6\textwidth]{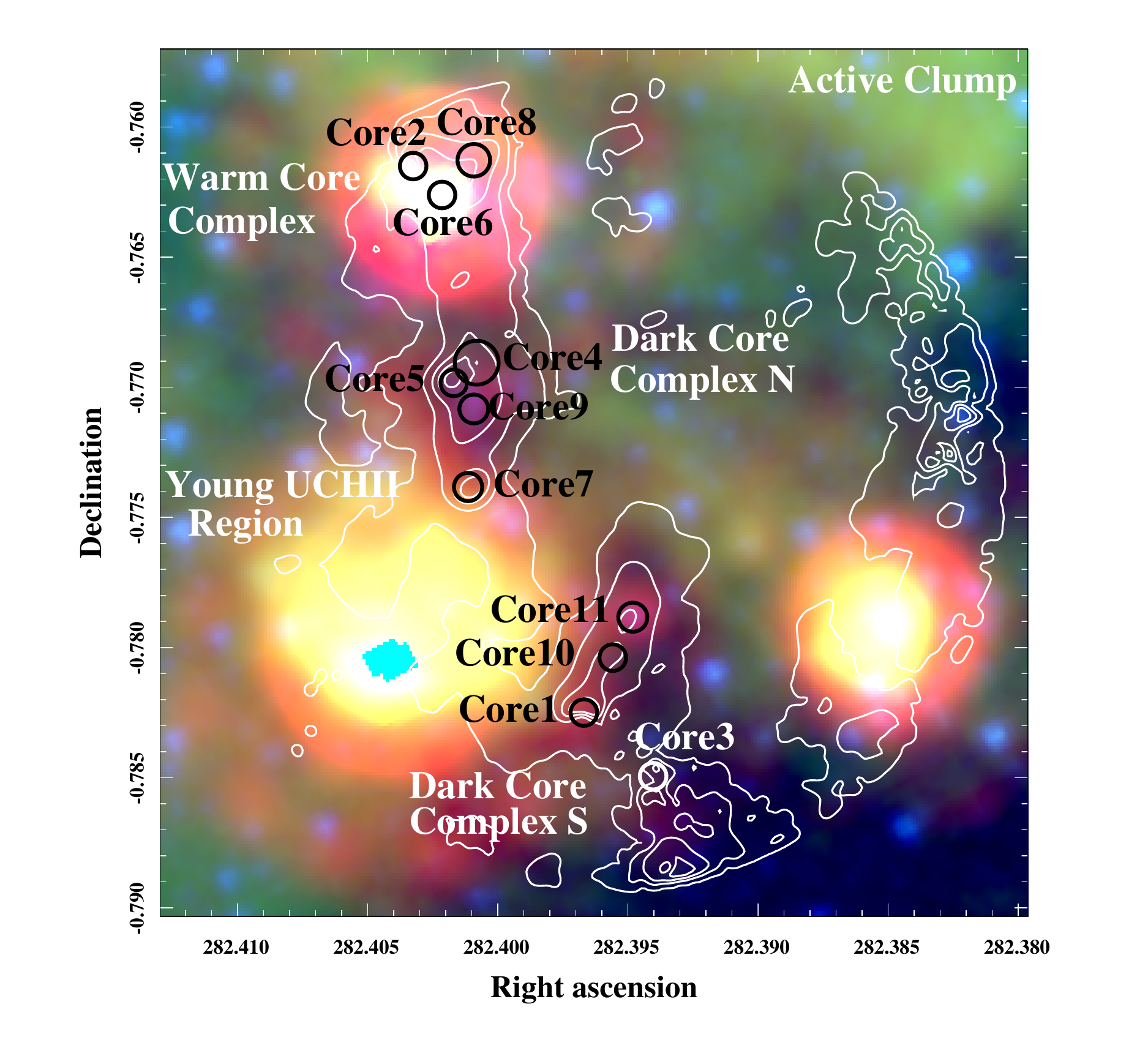}} \\
\subfigure{
\includegraphics[width=0.9\textwidth]{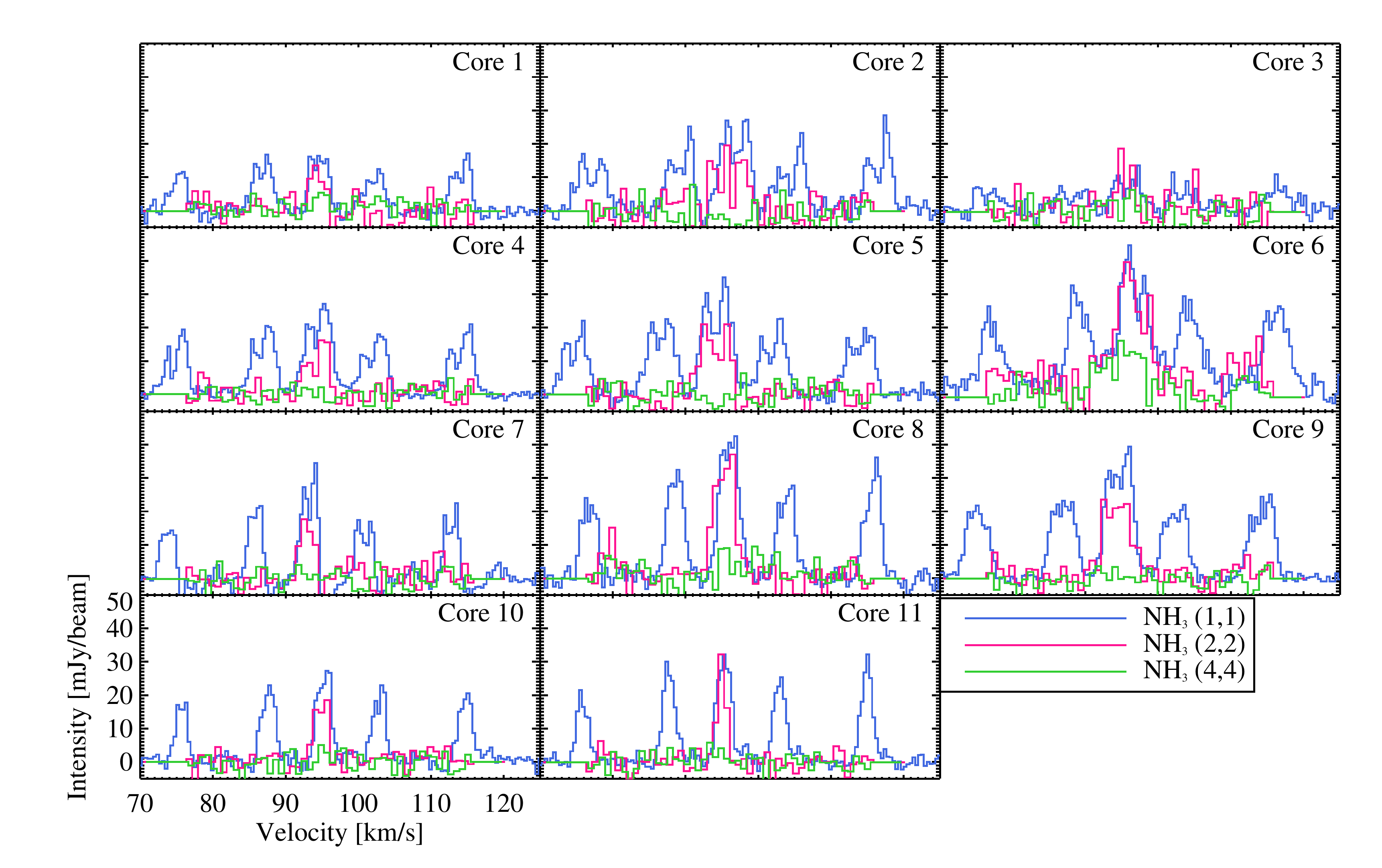}} \\
\caption{The \nh3 identified cores are labeled here for the active clump.  \textit{Top:} Three color mid-IR image of each clump (MIPSGAL, GLIMPSE, Red: 24 \micron, Green: 8 \micron, Blue: 4.5 \micron) with column density contours derived from \nh3 (six linearly spaced contours from 10$^{22}$ cm$^{-2}$ to 3$\times$10$^{23}$ cm$^{-2}$).  The core sizes and locations are plotted as circles with the effective radii of the 2-D Gaussian fits.  \textit{Bottom:} \nh3 (1,1), (2,2), and (4,4) spectra averaged over each core. The brightness temperature conversion for these data is about 9 mJy/beam per 1 K.} 
\label{fig:aspectra}
\end{figure*}

\begin{figure*}
  \centering
  \includegraphics[width=0.8\textwidth]{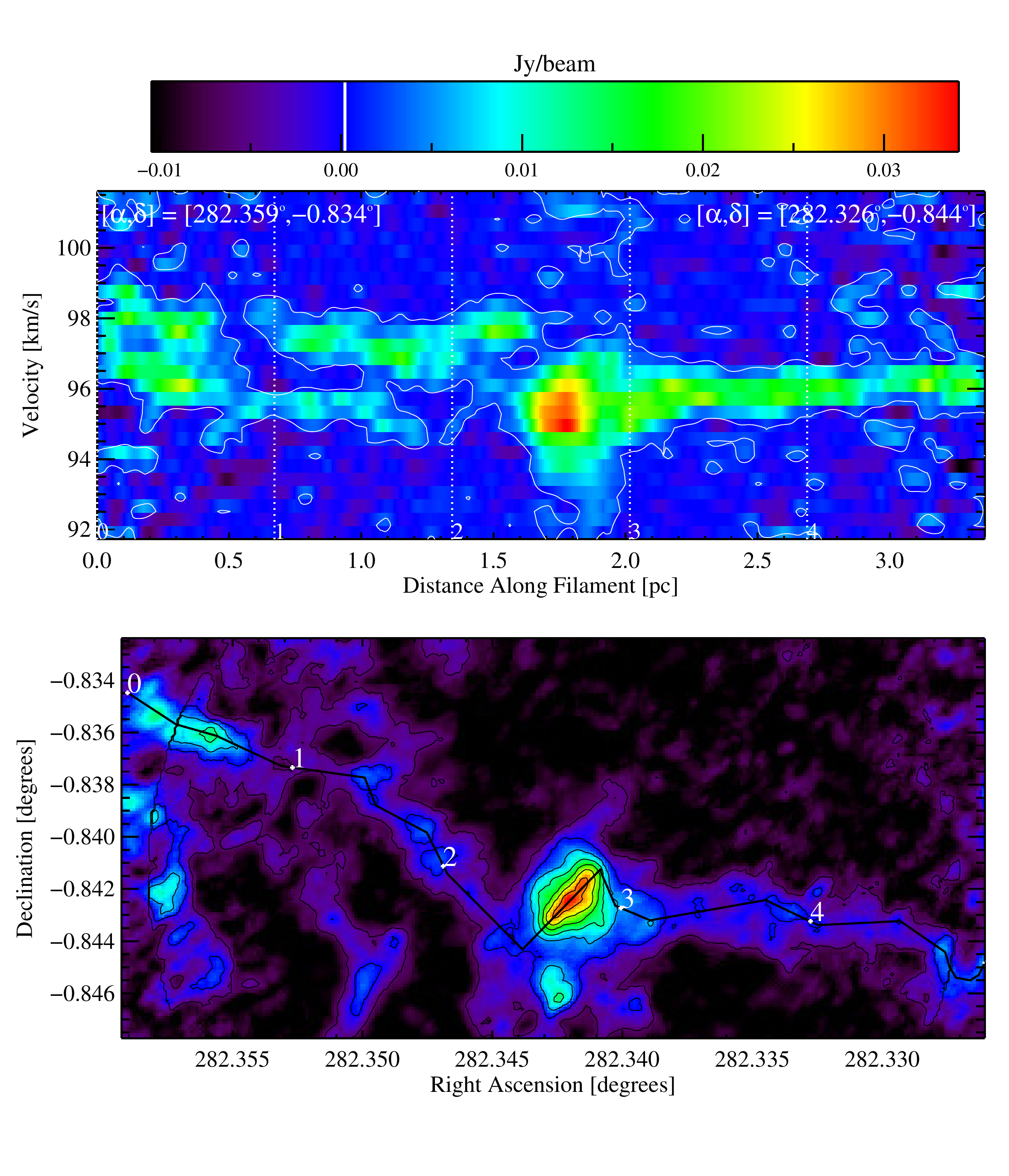}
  \caption{A position velocity diagram across one filamentary slice of the quiescent clump \nh3 (1,1) cube (shown as the black line in the bottom panel superimposed on the \nh3 (1,1) integrated intensity map, numbers from 0 to 4 corresponding to dotted vertical lines shown along the x axis of the top panel).  The low and high velocity components of the filament are connected in the easternmost portion, then diverge sharply at the central massive core.  As seen in Figure \ref{fig:velcomp}, the high velocity component spatially diverges west of the central massive core.}
  \label{fig:quiescent_pv}
\end{figure*}

\begin{figure*}
  \centering
  \subfigure{
    \includegraphics[width=0.55\textwidth]{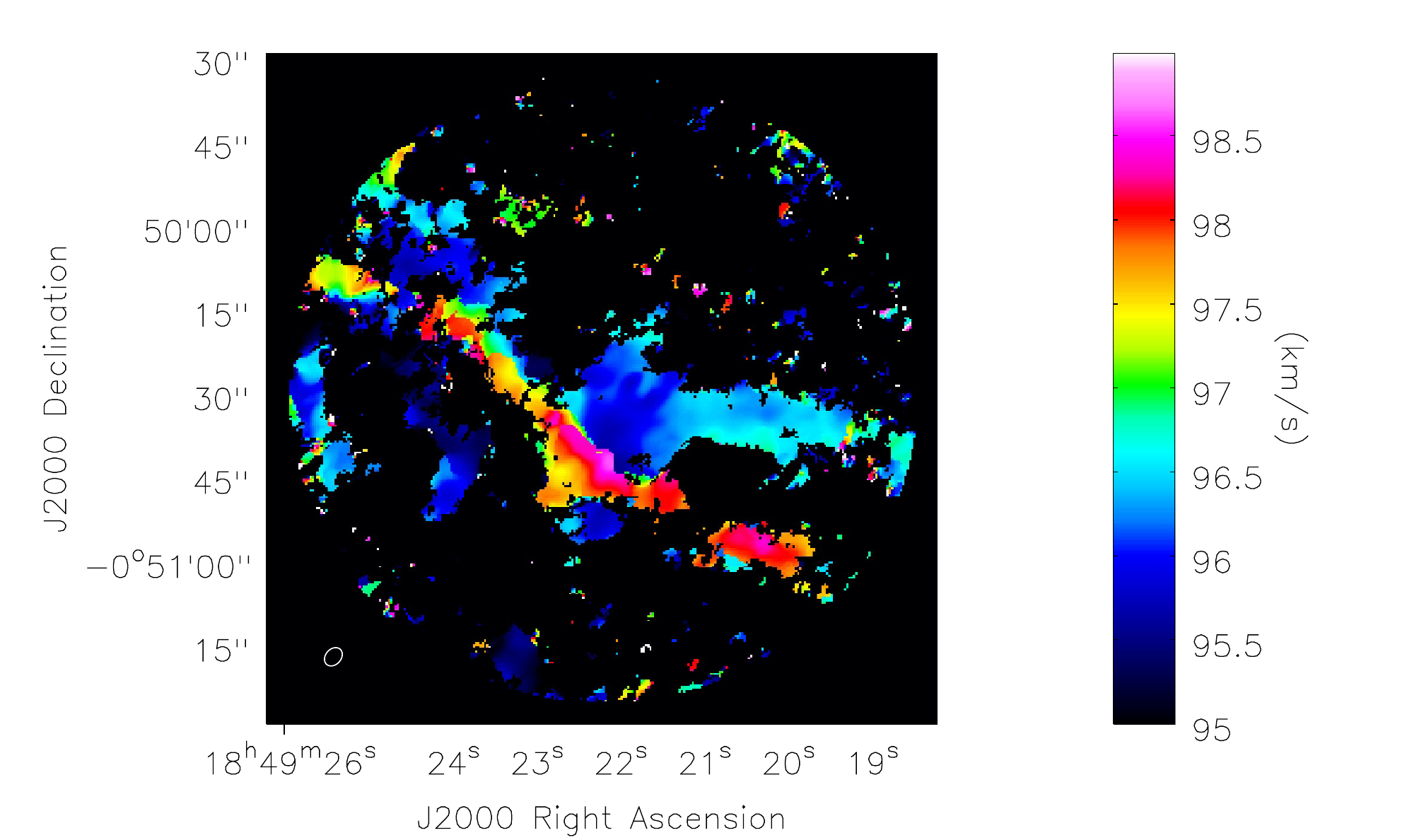}}
\hspace{-0.15in}
  \subfigure{
    \includegraphics[width=0.4\textwidth]{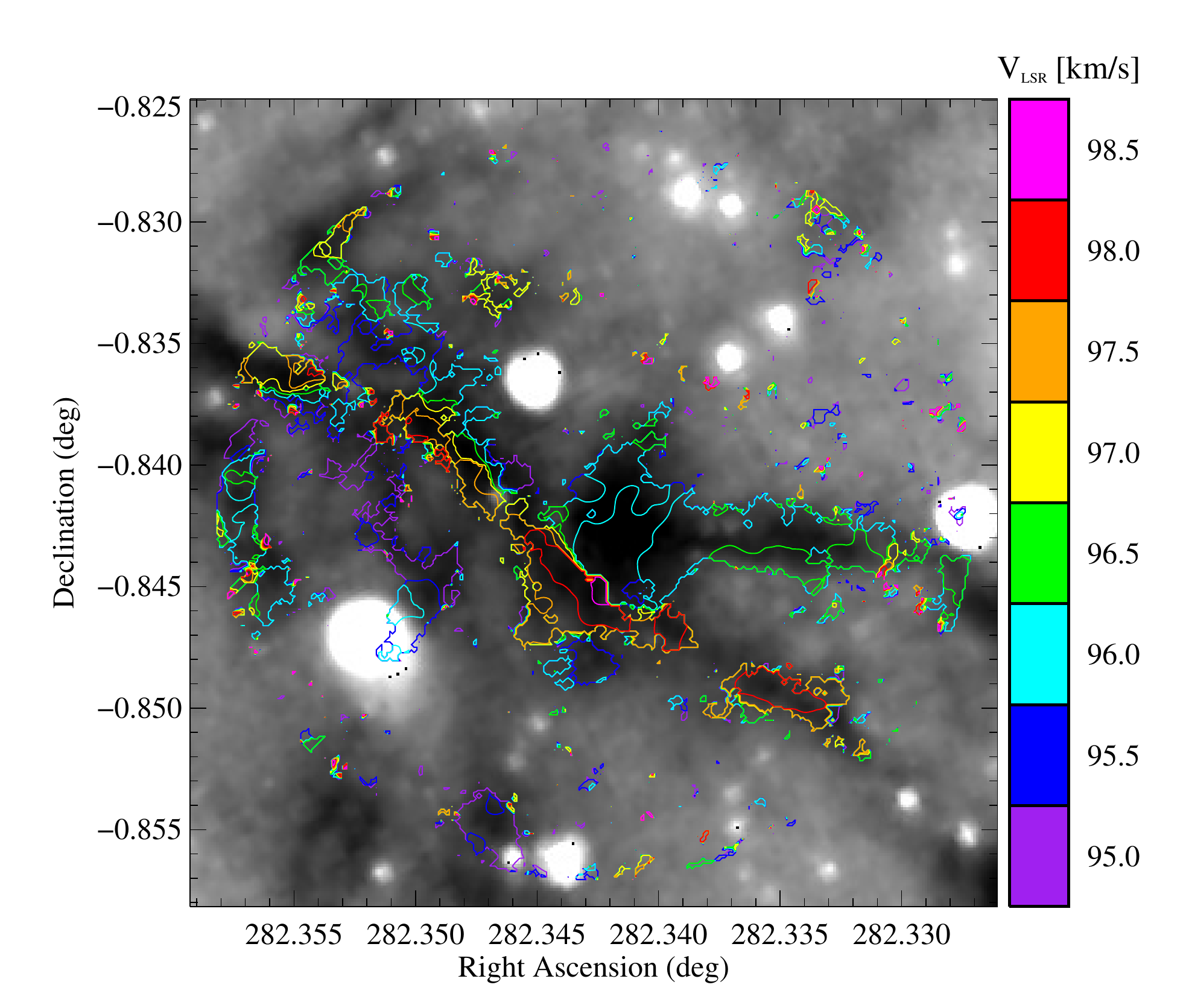}}
  \caption{Velocity field of the quiescent clump showing the two distinct filamentary components.  The low velocity component spans about 93 to 97.3 \kms~while the high velocity component spans about 97.3 to 101 \kms.  
\textit{Left:} The velocity field of the quiescent clump. \textit{Right:} The same
    velocity field shown as contours on a GLIMPSE 8 \micron~
    image. These images illustrate the complicated dynamics at work in
    a hub-filament system. The low velocity gradient across most of
    the filament may indicate its orientation in the plane of the sky,
    while the sharp velocity gradient near the hub center is
    indicative of flow dynamics near the site of early star formation.
}
\label{fig:velcomp}
\end{figure*}

\begin{figure*}
  \centering
  	\subfigure{
	\includegraphics[trim= 15mm 2mm 30mm 2mm, width=0.24\textwidth]{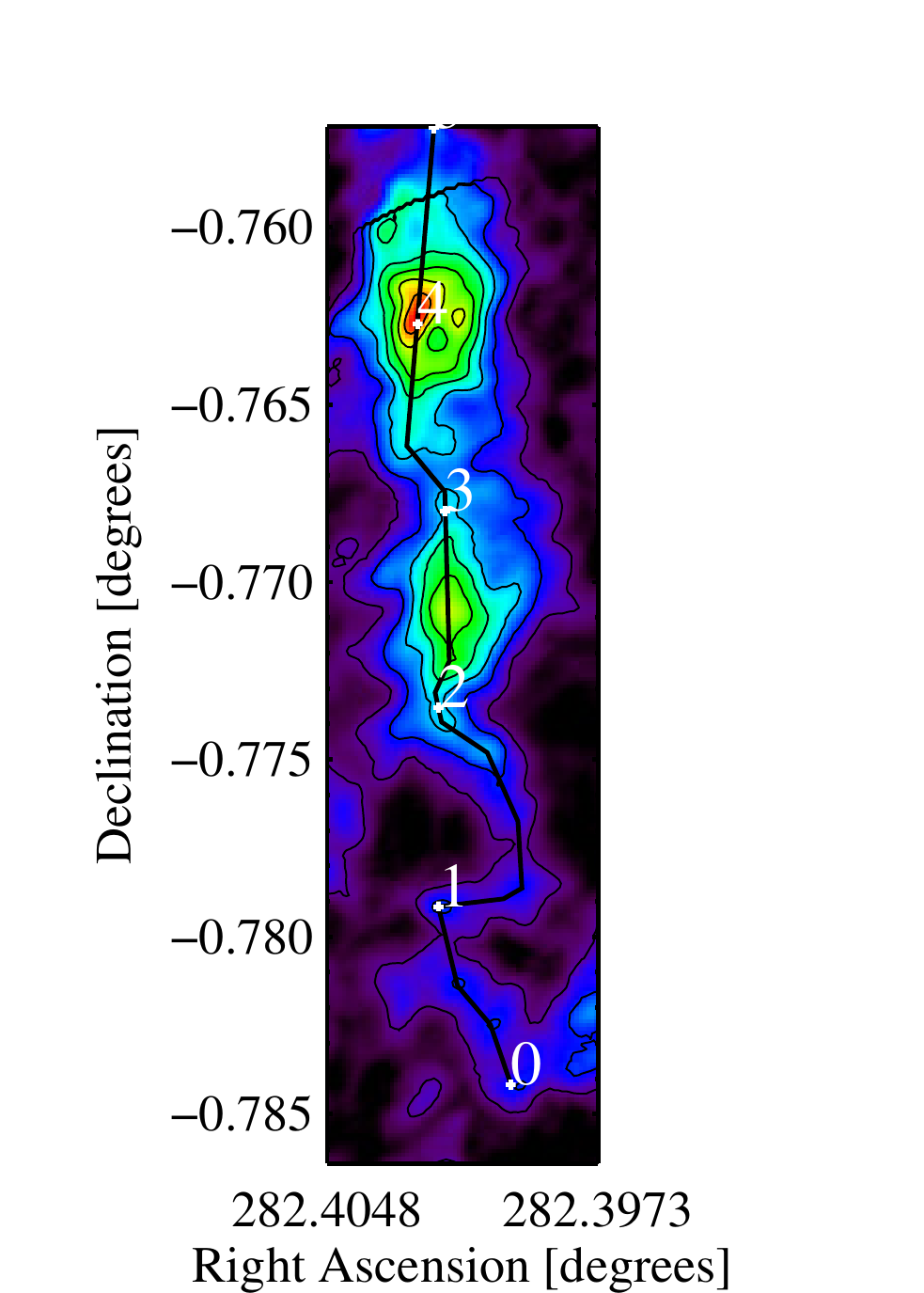}}
  \subfigure{
  \includegraphics[trim = 15mm 2mm 60mm 2mm, width=0.65\textwidth]{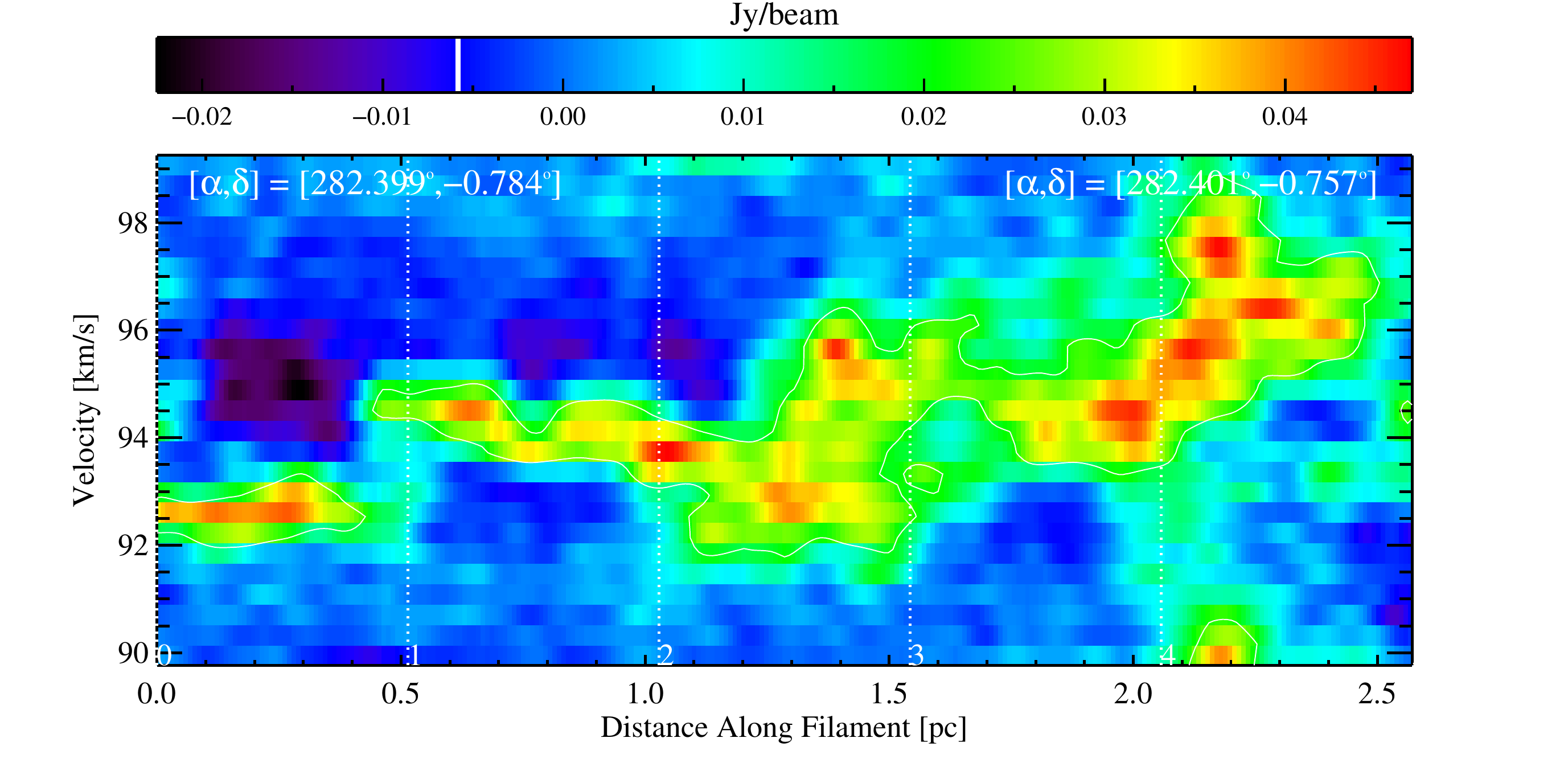}}
  \caption{A position velocity diagram across one filamentary slice of the active clump \nh3 (1,1) cube (shown as the black line in the left panel superimposed on the \nh3 (1,1) integrated intensity map, numbers from 0 to 4 corresponding to dotted vertical lines shown along the x axis of the right panel).  The PV diagram shows a complex structure, possibly consisting of massive cores, dense filaments, and bubbles driven by young stars.}
  \label{fig:active_pv}
\end{figure*}

\begin{figure*}
  \centering
     \includegraphics[width=1\textwidth]{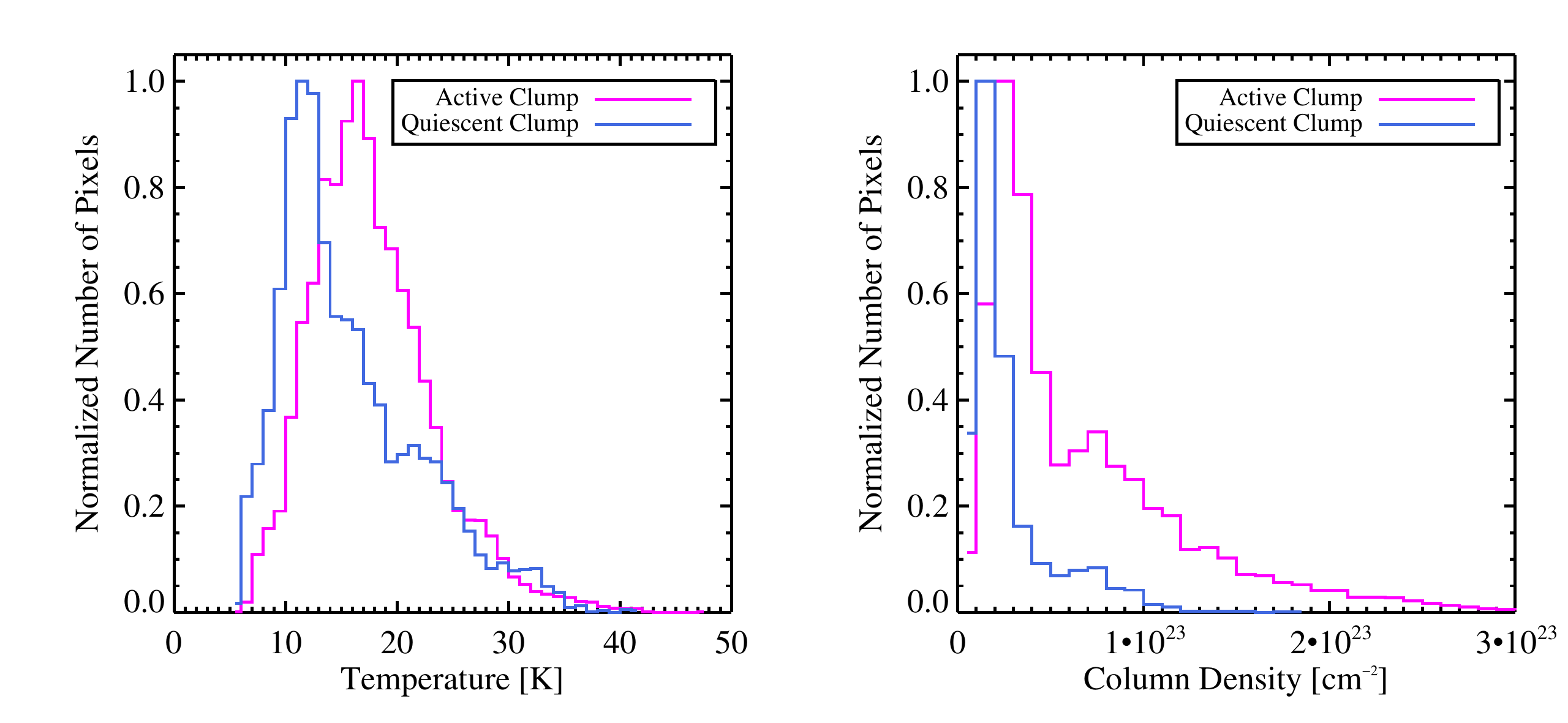}
  \caption{Histograms of the kinetic temperatures (\textit{left}) and column densities (\textit{right}) in the active and quiescent clumps.  Both distributions have a wide spread (standard deviation is $\sim$6 K in temperature for both clumps and $\sim$2$\times$10$^{22}$ cm$^{-2}$ in column density for the quiescent clump and $\sim$6$\times$10$^{22}$ cm$^{-2}$ for the active clump) but the median pixel value in the quiescent clump is  colder (14 K vs. 17 K) and lower column density (2$\times$10$^{22}$ cm$^{-2}$ vs. 5$\times$10$^{22}$ cm$^{-2}$) than the active clump.  The shapes of the column density histograms are affected by the image masking and spatial filter function, and small features in the histogram (such as the intriguing bump near 7.5 $\times$ 10$^{22}$ cm$^{-2}$) may be related to image artifacts.}
  \label{fig:temp_hist}
\end{figure*}


\begin{deluxetable}{llc}
\tabletypesize{\footnotesize}
\tablecaption{\textbf{Core Locations}\label{table-loc}}
\tablewidth{0pt}

\tablehead{
\colhead{} &
\colhead{Name} &
\colhead{($\alpha$, $\delta$)}  \\
\colhead{} &
\colhead{} & 
\colhead{(J2000)} 
}
\startdata

\multicolumn{2}{l}{Quiescent Clump}  \\
\cline{1-2}
  &   Core1        &  (18:49:22.0,-0:50:31.8)    \\
  &   Core2        &  (18:49:25.4,-0:50:08.6)   \\
  &   Core3        &  (18:49:21.1,-0:50:33.8)    \\
  &   Core4        &  (18:49:20.4,-0:50:33.4)     \\
  &   Core5        &  (18:49:20.1,-0:50:33.8)     \\
\multicolumn{2}{l}{Active Clump}   \\
\cline{1-2}
\multicolumn{2}{l}{Hot Complex}   \\
  &   Core2        &  (18:49:36.8,-0:45:41.4)  \\
  &   Core6        &  (18:49:36.5,-0:45:45.4)    \\
  &   Core8        &  (18:49:36.2,-0:45:40.6)  \\ 
 \multicolumn{2}{l}{Dark Complex N}    \\
  &   Core4        &  (18:49:36.2,-0:46:08.6)    \\
  &   Core5        &  (18:49:36.4,-0:46:11.4)  \\
  &   Core7        &  (18:49:36.3,-0:46:25.8)   \\
  &   Core9        &  (18:49:36.2,-0:46:15.0)   \\
\multicolumn{2}{l}{Dark Complex S}  \\
  &   Core1        &  (18:49:35.2,-0:46:57.0)  \\
  &   Core3        &  (18:49:34.6,-0:47:05.8)   \\
  &   Core10      &  (18:49:34.9,-0:46:49.4) \\
  &   Core11      &  (18:49:34.7,-0:46:43.8)  \\
 
 \enddata

\end{deluxetable}

\begin{deluxetable*}{llccccccccc}
\tabletypesize{\footnotesize}
\tablecaption{Core Complex and Filament Properties\tablenotemark{1} \label{table-ccprop}}
\tablewidth{0pt}
\tablehead{
\colhead{} &
\colhead{Name} & 
\colhead{Radius\tablenotemark{a} } &
\colhead{Avg. $\sigma$$_{v}$} &
\colhead{Mass\tablenotemark{b} } &
\colhead{M$_{\rm{MSF}}$(r)\tablenotemark{c} } & 
\colhead{Peak N(H$_{2}$)} &
\colhead{Peak n(H$_{2}$)\tablenotemark{d}} &
\colhead{Peak $\Sigma_{\rm{H}_{2}}$ } &
\colhead{T$_{\rm{mean}}$} &
\colhead{T$_{\rm{max}}$} \\
\colhead{} &
\colhead{} & 
\colhead{[pc]} &
 \colhead{[\kms]} &
\colhead{ [\Msun]} &
\colhead{ [\Msun]} &
\colhead{[cm$^{-2}$]} &
\colhead{[cm$^{-3}$]} &
\colhead{[g cm$^{-2}$]} &
\colhead{[K]} &
\colhead{[K]} \\
}

\startdata

\multicolumn{2}{l}{Quiescent Clump}  & &  &  &  &  &   &  & \\
\cline{1-2}
  &   Core1        &  $<$ 0.12 & 0.5  &  65  & 52 &  8.6 $\times$10$^{22}$  & $>$ 1.2 $\times$10$^{5}$  &   0.41   &  13  &  15  \\
  &   Core2        &  $<$ 0.07 & 0.2  &  35  & 25 &  1.0 $\times$10$^{23}$  & $>$ 2.4 $\times$10$^{5}$  &   0.49   &  20  &  24  \\
  &   Core3        &  $<$ 0.07 & 0.2  &  7  & 25 &     2 $\times$10$^{22}$  & $>$ 5 $\times$10$^{4}$  &   0.10  &  13  &  21  \\
  &   Core4        &  $<$ 0.07 & 0.2  &  6    & 25 &     2 $\times$10$^{22}$  & $>$ 4 $\times$10$^{4}$  &  0.08    &  15   &  27  \\
  &   Core5        &  $<$ 0.09 & 0.2  &  10  & 35 &   2 $\times$10$^{22}$  & $>$ 3 $\times$10$^{4}$  &    0.09   &  12   &  24  \\
  &  Filament     &  $<$ 0.08 & --    & 2500 &  &   Avg$\sim$2 $\times 10^{22}$  & $>$ 4 $\times$10$^{4}$  &   0.09   &  10-15   &  --  \\
\multicolumn{2}{l}{Active Clump}   &  &  &  &  &  &  &  \\
\cline{1-2}
\multicolumn{2}{l}{Hot Complex}  & &  &  &  &  &  &  &  \\
  &   Core2        &  $<$ 0.07 & 1.1 &  150  & 25 &  4.3 $\times$10$^{23}$  & $>$ 1.0 $\times$10$^{6}$  &   2.1   &  14 &  22  \\
  &   Core6        &  $<$ 0.07 & 1.4 &  76    & 25 &  2.2 $\times$10$^{23}$  & $>$ 5.2 $\times$10$^{5}$  &   1.1   &  34 &  42  \\
  &   Core8        &   $<$ 0.08 & 0.9 &  97    & 30 &  1.9 $\times$10$^{23}$  & $>$ 3.6 $\times$10$^{5}$  &   0.9  &  25 &  30  \\ 
 \multicolumn{2}{l}{Dark Complex N}  & &  &  &  &  &  &  &  \\
  &   Core4        &  $<$ 0.11 & 1.0  & 180  & 46 &  3.0 $\times$10$^{23}$  & $>$ 4.3 $\times$10$^{5}$  &   1.4   &  14 &  17  \\
  &   Core5        & $<$ 0.07 & 1.2  &  58   & 25 &  1.7 $\times$10$^{23}$  & $>$ 4.1 $\times$10$^{5}$  &   0.82  &  15 &  20  \\
  &   Core7        & $<$ 0.07 & 0.5  &  89   & 25 &  2.3 $\times$10$^{23}$  & $>$ 5.1 $\times$10$^{5}$  &   1.1   &  17 &  21  \\
  &   Core9        &  $<$ 0.08 & 1.2  & 53   & 30 &  1.3 $\times$10$^{23}$  & $>$ 2.8 $\times$10$^{5}$  &   0.63   &  16 &  18  \\
\multicolumn{2}{l}{Dark Complex S} & &  &  &  &  &  &  &  \\
  &   Core1        &  $<$ 0.07 & 0.5 & 210  & 25 &  6.2 $\times$10$^{23}$  & $>$ 1.5 $\times$10$^{6}$  &  2.9    &  14 &  22  \\
  &   Core3        &  $<$ 0.07 & 1.0 &  110   & 25 &  3.2 $\times$10$^{23}$  & $>$ 7.5 $\times$10$^{5}$  &  1.5   &  23 &  27  \\
  &   Core10      &  $<$ 0.07 & 0.6 &  34  & 25 &  1.0 $\times$10$^{23}$  & $>$ 2.4 $\times$10$^{5}$  &  0.48    &  18 &  21  \\
  &   Core11      &   $<$ 0.08 & 0.6 &  52  & 30 &  1.3 $\times$10$^{23}$  & $>$ 2.8 $\times$10$^{5}$  &   0.62   &  17 &  20  \\
  &  Filament     &   $<$ 0.14 & -- & 4800  &  &   Avg$\sim$5$\times 10^{22}$  & $>$ 6 $\times$10$^{4}$  &   0.23   &  15-20   &  --  \\

\enddata
\tablenotetext{1}{Typical statistical and systematic uncertainties are not included here, but are discussed in detail in \S \ref{sec:errors}.}
\tablenotetext{a}{The effective radius of the 2-D Gaussian Fit (a circle with this radius has the same area as the area under the Gaussian, r$_{eff}$ = $\sqrt{2}$$\sigma$).  The resolution limit is $\sim$0.07 pc so many of these are unresolved.  The source sizes are not de-convolved (and some sources are unresolved) and are therefore upper limits, the de-convolved source sizes are presented in \S \ref{sec:coreident}.  For the filament, the value reported is approximately the \textit{radius} of a cylindrical filament} 
\tablenotetext{b}{The mass is from the Gaussian fit to the \nh3-derived N(H$_{2}$) column density map, except in Quiescent Core 1 and Active Core 4, where the cores are resolved, so the mass is summed from the column density map over the core.}
\tablenotetext{c}{Mass threshold for forming massive stars within the given radius, from \citet{kau10}.}
\tablenotetext{d}{The peak N(H$_{2}$) divided by 2$\times$ the radius.  These densities are all lower limits since the radii are all upper limits.}

\end{deluxetable*}


\begin{deluxetable*}{llcccccc}
\tablecaption{Core Complex and Filament Star-Forming Activity \label{table-ccsf}}
\tablewidth{0pt}
\tablehead{
\colhead{} &
\colhead{Complex Name\tablenotemark{b}} & 
\colhead{8 \micron~signature} &
\colhead{24 \micron~source?} &
\colhead{Free-Free?\tablenotemark{a}} &
\colhead{6.7 GHz CH$_{3}$OH Maser?\tablenotemark{c}}  &
\colhead{Evolutionary Stage\tablenotemark{d}} \\
}
\startdata

\multicolumn{2}{l}{Quiescent Clump}  &  &  &  &  & \\
\cline{1-2}
  &  Core 1                       &  Dark  &  No  &  No  &  No  &  1  \\
  & Cores 2, 3, 4, and 5       &  Dark  &  No  &  No  &  No &  1  \\
  &  Filament                    &  Dark  &  No  &  No  &  No  &  1  \\
\multicolumn{2}{l}{Active Clump}  &  &  &  &  &  \\
\cline{1-2}
  &  Warm core Complex       & Bright  &  Yes  &  No  &  No   &  3  \\
  &  Dark Core Complex N & Dark  &  No    &  No  &  Yes\tablenotemark{c}  &  2  \\
  &  Dark Core Complex S & Dark  &  No    &  No  &  Yes\tablenotemark{c}  &  2  \\
  &  UCHII Region           &  Bright &  Yes  &  Yes &  No &   4  \\
  &  Filament           &  Dark  &  No  &  No  &  No &  1  \\
\enddata
\tablenotetext{a}{See \S \ref{sec:freefree}}
\tablenotetext{b}{Complexes are shown in Figure \ref{fig:aspectra}}
\tablenotetext{c}{\citet{pes05, szy02}.  The positional accuracy of this detection is about 30\arcsec~and the position overlaps both dark core complexes.}
\tablenotetext{d}{\citet{bat10}}

\end{deluxetable*}

\section{Results}
\label{sec:results}

\subsection{Errors}
\label{sec:errors}
We summarize here the statistical and model errors, as well as estimated systematic errors on derived quantities.  As discussed in \S \ref{sec:obsdet} and \ref{sec:reduction}, typical statistical errors on the \nh3 fluxes are about 6 mJy/beam.  This translates to typical model fit errors in the temperatures of 1-3 K and about 10\% ($\sigma$(log[N(\nh3)]) $\lesssim$ 0.05) in N(\nh3).  Given typical distance (20\%), opacity (100\%), and flux density (10\%) uncertainties, \citet{bat10}, using Monte Carlo simulations, found that typical systematic mass uncertainties are roughly a factor of two.  This does not include the unknown uncertainties associated with depletion, abundance variations, line blending, and varying excitation conditions in different \nh3 lines.  The systematic uncertainties in the temperature are also due to those unknown uncertainties which we are unable to quantify.  In summary, the masses are uncertain by about a factor of two and the column densities are slightly better (no uncertainty in distance).  

The temperatures are at best good to 1-3 K, but represent an average temperature of gas within the beam and not necessarily the underlying distribution.  The uncertainties in the radii are due to distance uncertainties and resolution effects.  The distance uncertainty gives about a 20\% uncertainty in the radius, while the resolution effect simply means that we are only sensitive to source sizes larger than our beam (about 0.07 pc radius) and true source sizes measured to be 0.07 pc are certainly smaller.

\subsection{The Model Fits}
\label{sec:modelfits}

The radiative transfer model fits are presented in Figures \ref{fig:coldlow_fits}, \ref{fig:coldhigh_fits} (quiescent clump), and \ref{fig:hot_fits} (active clump).  The velocity fields are shown as position velocity slices along the filaments in Figure \ref{fig:quiescent_pv} and \ref{fig:active_pv}.  There are two distinct kinematic components (referred to as the ``low", from 93 to 97.3 \kms, and ``high", from 97.3 to 101 \kms, velocity components) in the quiescent clump which were fit separately.  These components are depicted in Figure \ref{fig:velcomp}.  The ``low" and ``high" velocity components are clearly associated (spatial overlap and very close in velocity space) with the strongest overlap region being in the Northeast end of the filament (near $\alpha$ = 282.355, $\delta$ = -0.837, Core 2). 
These kinematically distinct interacting filaments bear morphological similarity to W33A as presented by \citet{gal10}.

The active clump also consists of multiple components.  However, these were heavily blended and therefore were unable to be fit separately.  Sharp gradients in the fits occurred when the routine	jumped between the two components because noise made them comparably significant.  To create a smooth, averaged gradient across the components, the model fits were convolved with the observed beam ($\sim$3\arcsec).  This reduces the effective resolution slightly and overestimates the size scale of the different cores along the line of sight.

\subsubsection{Core Identification}
\label{sec:coreident}
Cores were identified in the column density maps by fitting a 2-D Gaussian to the brightest point in the map, subtracting it off, and re-iterating \citep[similar to][]{bro09, rat06}.  All the core fits were inspected by eye.  The cores are plotted in Figures \ref{fig:qspectra} and \ref{fig:aspectra} with their effective radii (a circle with this radius has the same area as the area under the Gaussian, r$_{eff}$ = $\sqrt{2}$$\sigma$).  The core parameters reported in Table \ref{table-ccprop} (central positions reported in Table \ref{table-loc}) are derived within this effective radius.  This method of identifying cores does not provide a complete sample and is meant to be representative of the brightest cores.  The core sizes reported in Table \ref{table-ccprop} and throughout are not de-convolved.  The sizes of cores with reported radii of 0.07 pc are unresolved and their sizes are only constrained to be smaller than the 0.07 pc resolution limit.  The de-convolved source radii of 0.08 pc, 0.09 pc, 0.11 pc, and 0.12 pc sources are 0.04 pc, 0.06 pc, 0.08 pc, and 0.10 pc, respectively.

\subsubsection{The Temperature Structure of the Clumps}
The kinetic temperature histogram is shown in Figure \ref{fig:temp_hist}, the maps are shown in Figures \ref{fig:coldlow_fits}, \ref{fig:coldhigh_fits}, and \ref{fig:hot_fits} and the region properties are summarized in Tables \ref{table-ccprop} and \ref{table-ccsf}, while the sub-regions discussed are labeled in Figures \ref{fig:qspectra} and \ref{fig:aspectra}.  Both clumps range from about 8 to 35 K.  The quiescent clump exhibits a smooth spatial distribution in temperature in the interior of the filament from about 8 to 15 K.  The temperature decreases towards the interior of the filament from about 20 K to 12 K. Interestingly, the main core (Core 1) of the quiescent clump shows no gradient in temperature and is approximately 13 K.  The high velocity kinematic component of the quiescent clump shows a clumpier and slightly warmer ($\sim$ 15 K) temperature structure.  Core 2 is approximately the overlap region of the two velocity components making the parameter estimation more difficult, but the derived kinetic temperature is about 20 K.  Cores 3, 4, and 5 are all colder at about 12-15 K, but their column density structures do not seem to be well approximated by a simple 2-D Gaussian.

The temperature structure of the active clump is more varied than the quiescent clump.  The temperature ranges from about 10 K in the coldest portions to over 40 K in the warmest core (Core 6).  The temperature structure appears to be fairly smooth and cold (10 to 20 K) over the majority of the active filament which is only disrupted by the warm pockets (30-40 K) of star formation.  The active clump has both warm (40 K)  and cold (12-15 K) cores.  The warm core complex in the north of the filament has a smooth, warm temperature of about 20 K with gradual increases to about 40 K toward Core 2, but not directly peaking at the core center, perhaps indicating a newly formed star providing an internal heating source (at the temperature peak) and a pre-stellar dense core (at the column density peak).  Both of the dark core complexes (N and S) have smooth, cold temperature fields of about 12-15 K.  Most of the cores in the dark core complexes show no gradient in temperature and are roughly the same temperature as the filament at about 15 K.  However, the southernmost core (Core 7) in the dark core complex N shows a decrease in temperature toward its center, from $\sim$ 20 K to 15 K. 

\subsubsection{The Column Density Structure of the Clumps}
\begin{figure}
  \centering
     \includegraphics[width=0.5\textwidth]{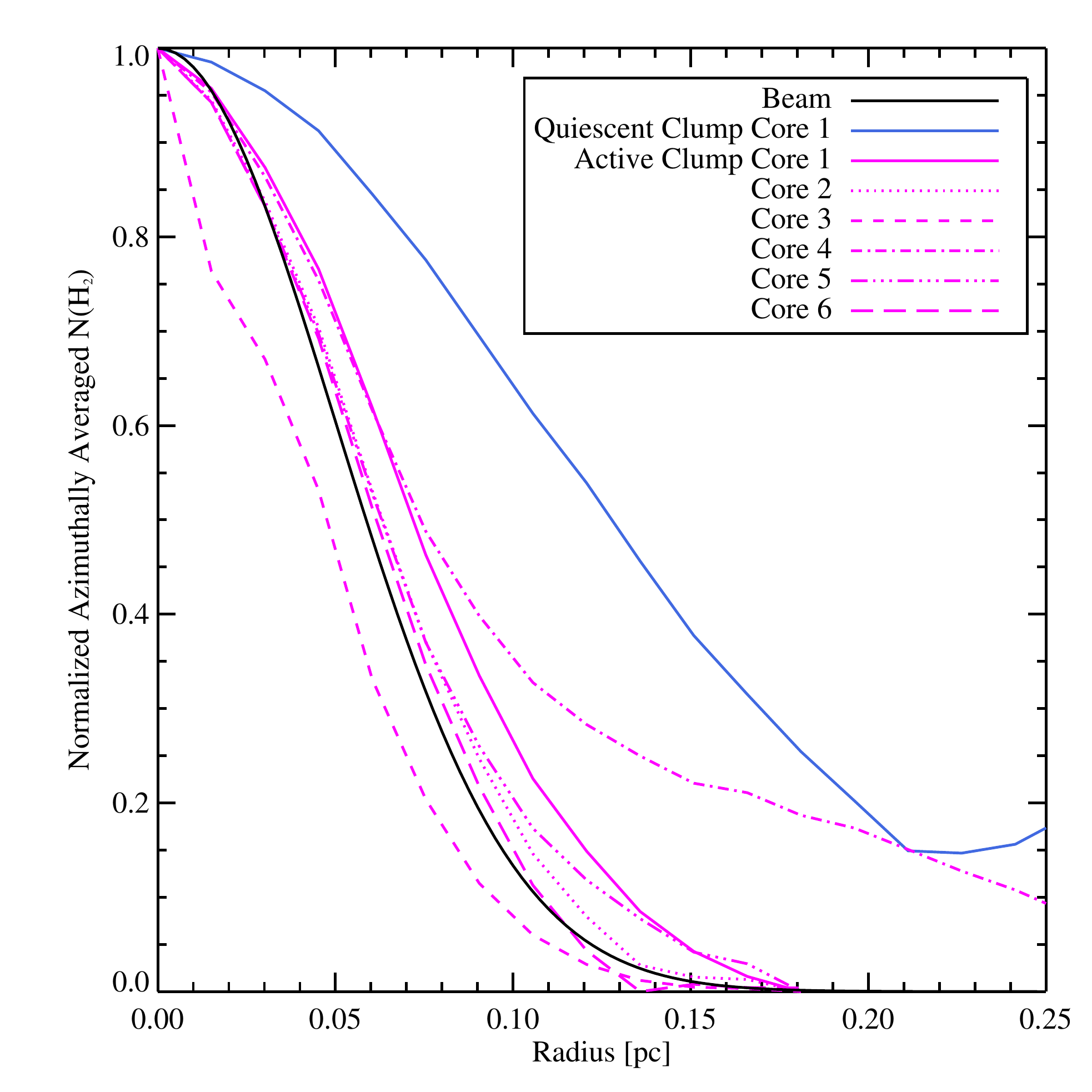}
  \caption{The normalized azimuthally averaged N(H$_{2}$) radial profiles of some of the brightest cores as a function of radius in pc.  Most are just barely resolved and mimic the beam profile (shown in black), while Quiescent Core 1 (blue solid line) is clearly much more extended.  Core 3 has unsampled points around it, hence its having a lower azimuthally averaged profile than the beam.
  }
  \label{fig:radial}
\end{figure}

The column density histogram for both clumps is shown in Figure \ref{fig:temp_hist}, the maps are shown in Figures \ref{fig:coldlow_fits}, \ref{fig:coldhigh_fits}, and region properties are summarized in Tables \ref{table-ccprop} and \ref{table-ccsf}, while the sub-regions discussed are labeled in Figures \ref{fig:qspectra} and \ref{fig:aspectra}.  The azimuthally averaged N(H$_{2}$) radial profiles are shown in Figure \ref{fig:radial}.  The active clump has an overall higher column density, with a peak column density of about 3.3 $\times$ 10$^{23}$ cm$^{-2}$ compared to about 1.1 $\times$ 10$^{23}$ cm$^{-2}$ in the quiescent clump.

\textit{The Active Clump: }There are three main clusters of cores in the active filament, a northern warm core complex (IR-bright, about 35-40 K) and two dark core complexes, North and South (IR-dark and cold, about 15 K) in the middle and bottom of the filament.  The young UCHII region (see \S \ref{sec:freefree}) seen as a bubble in \nh3 could indicate a previous core complex or a more massive core complex at the same age.  In the active clump, the cores are characterized by column densities of 2-3 $\times$ 10$^{23}$ cm$^{-2}$ or surface densities of 1 g cm$^{-2}$ ($\sim$ 2.1 $\times$ 10$^{23}$ cm$^{-2}$), just above the theoretical limit for forming massive stars \citep{kru08}.  The filament in the active clump ranges from about 3 $\times$ 10$^{22}$ cm$^{-2}$ to 1 $\times$ 10$^{23}$ cm$^{-2}$. 

The young UCHII Region and the three core complexes in the active clump along the filament are separated by about 30\arcsec~(0.8 $\pm$ 0.1 pc).  The cores themselves remain nearly unresolved (at 0.1 pc) but their separations (average separation to nearest neighbor within each complex) are about 6\arcsec~(0.16 $\pm$ 0.04 pc) with two larger (11\arcsec~and 13\arcsec, $\sim$0.3 pc) separations.

\textit{The Quiescent Clump: }  
The quiescent clump has one large core (Core 1, about 9\arcsec~or 0.24 pc diameter) at the center with a few more small, low significance cores throughout.  The central core has a column density of about 8.6 $\times$ 10$^{22}$ cm$^{-2}$.  This core at present is just below the predicted 1 g cm$^{-2}$ theoretical threshold for forming massive stars \citep[][]{kru08}, but is still very extended (see Figure \ref{fig:radial}).  If this core condensed by a factor of $\sqrt{2}$ in radius to 6\arcsec~it could reach this predicted threshold, or higher since a significant amount of large scale structure is missed in the interferometer observations.  As seen Table \ref{table-ccprop}, Core 1 is above the observational cutoff for massive star formation determined in \citet{kau10}, so it is plausible that the quiescent clump will also form massive stars.  Moreover, Core 1 shows a remarkable lack of substructure and a very smooth radial profile.  Possibly, the cold core may condense further through gravitational contraction and fragment to produce a core complex similar to that seen in the active clump.  Alternatively, given typical mass uncertainties of about a factor of two, this clump may not form massive stars at all, and therefore not be a Stage 1 clump. Because of the cold temperatures and low velocity dispersion, it is possible that this core will fragment and form only low-intermediate mass stars. However, the lack of substructure, smooth radial profile, and high surface density make Core 1 a good candidate precursor to massive star formation.

The overall mass (about 3000 \Msun) of the entire quiescent clump is about 2 to 3 times less than the active clump, and so would need to accrete significant mass were it to mirror the active clump, a scenario suggested to occur in massive star forming regions \citep[e.g.,][]{lon11}.  The secondary, low significance cores have column densities of about 2-4 $\times$ 10$^{22}$ cm$^{-2}$ and sizes near the resolution limit at 4\arcsec~(0.1 pc, see Table \ref{table-ccprop}) and do not seem well characterized by a simple 2-D Gaussian.  The core separations range from 5\arcsec~to about 10\arcsec~to about 1\arcmin~(0.1 pc to 0.3 pc to 1.5 pc).

\subsection{Fragmentation}
\label{sec:frag}
The young UCHII Region and the three core complexes in the active clump along the filament are each separated by about 0.8 ($\pm$ 0.1) pc.  The cores themselves remain nearly unresolved but their separations (average separation to nearest neighbor within each complex) are about 0.16 ($\pm$ 0.04) pc.  The Jeans length in the active clump given typical conditions (n = 6$\times$10$^{5}$ cm$^{-3}$, T = 40 K and T = 15 K, respectively) is 0.03 pc (for 40 K, 0.02 pc for 15 K).  The cores in the quiescent clump are separated by 0.1 to 1.5 pc, while a typical Jeans length is about 0.04 pc (n = 9$\times$10$^{4}$ cm$^{-3}$, T = 10 K).  The thermal Jeans length is well below our resolution limit.  The fragmentation scale we observe in the active clump likely indicates that turbulence governs the fragmentation length scale \citep[as in e.g.,][]{wan14, zha11, li12}.  

\subsection{Virial Parameter}
\label{sec:virial}
We calculate a virial mass and virial parameter for each core using the average velocity dispersion in each core.  The velocity dispersion is calculated as a free parameter in the radiative transfer model of all the \nh3 transitions (discussed in \S \ref{sec:nh3model}).  The calculation for the virial mass assumes that the core is bound and in virial equilibrium.  If the derived virial mass is less than the measured core mass ($\alpha_{{\rm vir}}$ $<$ 1), the core is sub-virial and may be collapsing.  Alternatively, the core may be in equilibrium and contain additional pressure support from, e.g. magnetic fields.  For a core with a density profile of $\rho$(r) $\propto$ r$^{-p}$, the virial mass is given by 
\begin{equation}
{\rm M_{vir}} = 3 \bigg[\frac{5-2p}{3-p} \bigg] \frac{{\rm R}\sigma^{2}}{{\rm G}}
\end{equation}
where $\sigma$ is the average line-of-sight velocity dispersion, R is the core radius, and G is the gravitational constant.  We adopt a value of p = 1.8 for the radial profile index, which was the mean in a sample of 31 massive star forming regions by \citet{mue02}.  We do not make any correction for ellipticity as the cores were all well approximated by circular 2-D Gaussians.  The expression for the virial mass reduces to 
\begin{equation}
{\rm M_{vir}} = 147 \bigg(\frac{{\rm R}}{1 {\rm pc}}\bigg) \bigg( \frac{\Delta{\rm v}_{{\rm fwhm}}}{1~{\rm km~s^{-1}}} \bigg)^{2}  {\rm M}_{\odot}.
\end{equation}
The virial masses and parameters are reported in Table \ref{table-virial}.  With the exception of Active Clump Cores 5, 6 and 9, all the cores are sub-virial, implying that they are unstable against collapse.  In other words, the thermal plus non-thermal pressure support as indicated by the \nh3 linewidths is insufficient to support most of the cores against collapse.  Even considering the systematic uncertainties of the mass estimates (see \S \ref{sec:errors}, about a factor of two), in many cases, the virial parameters are still constrained to be less than one.  This is similar to the results found in \citet{kau13b} where they compare virial parameters across a large sample of clouds and high mass star forming regions, finding low virial parameters, particularly toward the high mass star forming regions.  Finally, we note that \citet{bea13} find that projection effects alone in molecular clouds can lead to a factor of two uncertainty in the derived virial parameter using CO, however, our \nh3 dense gas measurements should be slightly better.


\begin{deluxetable}{llcccccc}
\tabletypesize{\footnotesize}

\tablecaption{\textbf{Virial Parameters}} \label{table-virial}
\tablewidth{0pt}


\tablehead{
 \colhead{} &
\colhead{Name} & 
\colhead{Radius$^{a}$}  &
\colhead{Mass$^{b}$}  &
\colhead{M$_{ {\rm vir}}$$^{c}$} &
\colhead{$\alpha_{ {\rm vir}}$$^{d}$} &
\colhead{Avg. $\sigma$$_{v}$} \\
\colhead{} &
\colhead{} & 
\colhead{[pc]} &
 \colhead{[\Msun]} &
\colhead{[\Msun]} &
\colhead{} &  
 \colhead{[\kms]}  \\ 
}

\startdata

\multicolumn{2}{l}{Quiescent Clump}  &   &  &  &  \\
\cline{1-2}
  &   Core1         & $<$ 0.12  &  65  & $<$ 25 &  $<$ 0.4 & 0.5   \\
  &   Core2        & $<$  0.07  &  35  & $<$ 2  &  $<$ 0.1 &  0.2    \\
  &   Core3        & $<$  0.07  &  7  & $<$ 2  &  $<$ 0.3  &  0.2    \\
  &   Core4        & $<$  0.07  &  6    & $<$ 2  &  $<$ 0.3 &  0.2    \\
  &   Core5        & $<$  0.09  &  10  & $<$ 4  &  $<$ 0.4 &  0.2     \\
\multicolumn{2}{l}{Active Clump}  &  &  &  &    \\
\cline{1-2}
\multicolumn{2}{l}{Hot Complex}  &  & &  &   \\
  &   Core2        & $<$  0.07 &  150  & $<$ 68  &  $<$ 0.5 &  1.1  \\
  &   Core6        & $<$  0.07 &  76    & $<$ 108 &  $<$ 1.4 &  1.4 \\
  &   Core8        &$<$   0.08 &  97    & $<$ 58 &  $<$ 0.6  & 0.9   \\ 
  \multicolumn{2}{l}{Dark Complex N}  &  &  &  &    \\
  &   Core4        & $<$  0.11  & 180  & $<$ 99 &  $<$ 0.6 & 1.0  \\
  &   Core5        & $<$  0.07  &  58   & $<$ 79 &  $<$ 1.4 &  1.2   \\
  &   Core7        & $<$  0.07  &  89   & $<$ 17 &  $<$ 0.2 & 0.5   \\
  &   Core9        & $<$  0.08  &  53   & $<$ 85 &  $<$ 1.6 &  1.2   \\
\multicolumn{2}{l}{Dark Complex S}  &  &  &  &    \\
  &   Core1        & $<$   0.07 & 210  & $<$ 16 &  $<$ 0.1 & 0.5    \\
  &   Core3        & $<$   0.07 & 110  & $<$ 61 &  $<$ 0.6 & 1.0  \\
  &   Core10      & $<$   0.07 &  34  & $<$ 21 &  $<$ 0.6 & 0.6   \\
  &   Core11      & $<$   0.08 &  52  & $<$ 22 &  $<$ 0.4 & 0.6   \\

\enddata
\tablenotetext{a}{The effective radius of the 2-D Gaussian Fit (a circle with this radius has the same area as the area under the Gaussian, r$_{eff}$ = $\sqrt{2}$$\sigma$).  The resolution limit is $\sim$0.07 pc so many of these are unresolved.  The source sizes are not de-convolved (and many are unresolved) and are therefore upper limits, the de-convolved source sizes are presented in \S \ref{sec:coreident}.  For the filament, the value reported is approximately the radius of the cylindrical filament.}
\tablenotetext{b}{The mass from the Gaussian fits to the \nh3 column density map, assuming \abnh3 = 4.6 $\times$ 10$^{-8}$.}
\tablenotetext{c}{Virial Mass as calculated in \S \ref{sec:virial}, these are upper limits as the radius is not de-convolved and some sources are unresolved.}
\tablenotetext{d}{$\alpha_{{\rm vir}}$ = M$_{{\rm vir}}$ / M, a value less than 1 implies that the core is unstable against collapse or has additional (non-thermal or turbulent) pressure support.  These are upper limits as the radius is not de-convolved and some sources are unresolved.}

\end{deluxetable}

\subsection{Evolutionary Stages}
\label{sec:ev}
The quiescent and active clumps observed in this IRDC show the full range of evolutionary stages from cold, dark, and quiescent (Stage 1) up to a young UCHII Region (Stage 4, see Table \ref{table-ccsf}).  We characterize the evolutionary stage of the core complexes in each clump according to the sequence presented in \citet{bat10} \citep[see also, e.g.,][]{cha09, pur09}.  This evolutionary sequence was derived by combing a physically motivated progression of massive star formation with common observational signatures.  We observe cold, quiescent IRDCs which often appear massive enough to form stars, but show no signs of active star formation \citep[e.g.][]{wil12}.  These regions are the best candidates for the pre-cursors of massive star formation and are therefore assigned as Stage 1.  Stage 2 sources remain infrared-dark but show early signs of massive star formation as signatures of shocks in outflows, like CH$_{3}$OH masers or Extended Green Objects \citep[EGOs;][]{cyg08, cha09}.  Stage 3 sources show signs of dust heating due to the energy released by gravitational contraction or the nuclear energy after a star has turned on in the form of mid-IR emission.  Stage 3 sources may include young UCHII regions which are still accreting from their natal dust cocoon while it is mostly unaffected by the ionizing radiation.  In Stage 4, the UCHII region has blown out much of the stellar cocoon and is observed as a bright mid-IR source, but with a diminishing amount of cold, dense gas.  This observational sequence bears much similarity to others \citep[e.g.][]{cha09, pur09} and remains to be tested.  \citet{bat11} demonstrated that temperatures increase systematically as a function of these star formation tracers which may indicate a positive evolutionary trend.  Factors which are not sequential but may influence the observational signatures are the core/outflow orientation, the clump mass, magnetic fields, and external pressure.  These evolutionary stages provide a guide for expected systematic trends in massive star formation not a rigorous rule book.

In doing this analysis we make the pivotal assumption that each core complex has a similar eventual fate and that differences between the complexes can be attributed to evolutionary stage alone.  While this crucial assumption has plausible reasoning (similar typical core masses, sizes, and separations, and the overall clump column density and temperature structure are very similar), it is inherently uncertain and it is important to take the analysis which follows with some degree of skepticism.  
In particular, differences which we here attribute to evolutionary stage, such as central density and temperature, could be due to inherent differences in the cores and their environments.  The following analysis rests on the assumption that we are sampling only a difference in evolutionary stage, but the regions may differ in other crucial ways.  While a large sample will improve the number statistics of any trend, it will also introduce additional variables through the range of environments, abundances, and distances sampled, while our cores span multiple evolutionary stages within a single cloud, much less hindered by those variations.

The quiescent clump shows little to no signs of active star formation (see \S \ref{sec:selection}, no 6.7 GHz CH$_{3}$OH maser, no significant 8 \micron~ emission or free-free emission, but there is a possible association with a faint 24 \micron~point source).  The active clump, however, shows a range of star formation activity consistent with the evolutionary sequence presented in \citet{bat10}.

The young UCHII region in the active clump is the most evolved, categorized to be in Stage 4 \citep{bat10}, followed by the Stage 3, 8 \micron~bright, warm (but lacking significant free-free emission) warm core complex in the north, while the two Stage 2, cold 8 \micron~dark core complexes seem to be the youngest.  The warm core complex and dark core complexes exhibit very similar properties in their density structure, both with column densities above the predicted threshold of 1 g cm$^{-2}$ for forming massive stars \citep{kru08}.  The warm core complex shows signs of a forming massive star, while the dark core complexes appear to be in a pre-stellar state.  The core complexes observed in this IRDC show good agreement with the evolutionary sequence from \citet{bat10} \citep[with the modification of the 24 \micron~phase from][]{bat11}, spanning the range from Stage 1 to Stage 4 with the derived temperatures increasing from the youngest to the most evolved phase.

Given the similar column density structure of the dark core complexes and the warm core complex (similar peak column densities, source sizes, separations, and total mass), the inference that these complexes represent similar regions in different stages of evolution seems reasonable.  One possible explanation is that the dark core complex may have fragmented below our resolution limit and is already forming low mass stars.  The structure of the quiescent clump is much more extended than any of the active clump core complexes.  If it represents an earlier evolutionary stage, then the clump needs to condense and fragment into dense cores.  
If the clump is undergoing global collapse or ongoing accretion, the peak central densities will rise with time.  An alternative hypothesis might be that the quiescent clump has or will fragment below our resolution limit and form low-mass stars.  A larger scale study would be required to more definitively address the evolution of core complexes and whether or not we are truly seeing an evolutionary sequence in high-mass star-forming regions, from quiescent and extended, to condensed and densely fragmented, finally to young, warm cores.  However, a larger study also introduces additional variables (varying distances, abundances, environments) affecting an inference of evolutionary trends.
Observations of a statistically significant sample of such core complexes would also place constraints on the timescale of core formation, their properties, and how long they survive in a pre-stellar state.  The observations presented here and in other works in the literature \citep[e.g.,][]{zha11, pil11, beu02} suggest that the youngest, coldest cores may be more extended, and gradually accrete material over time and then fragment into smaller cores.

\subsection{A Young HII Region in the Active Clump}
\label{sec:freefree}

The quiescent clump shows no sign of radio continuum emission in either our VLA maps (3 $\sigma$ continuum sensitivity of 0.6 mJy) or in the MAGPIS maps at 6 and 20 cm \citep{whi05, hel06}.  However, one continuum source is detected at ($\alpha$, $\delta$) = (282.404\deg, -0.779\deg), G$32.03 +0.05$, in the active clump and appears to be interacting with the surrounding dense \nh3 gas.  G$32.03+0.05$ is an extended source and we extracted peak and integrated flux densities (using the total of a 2-D Gaussian fit to the data) from the \nh3 data at 1.26 cm and from various epochs of the MAGPIS data at 6 and 20 cm \citep{whi05, hel06}, see also the newer CORNISH survey \citep{pur13, hoa12}.  Since this source is extended, the fluxes derived are highly sensitive to the largest angular scale of the observation.  
\citet{and11} point out that MAGPIS source fluxes are often lower limits as there is a good deal of diffuse emission missed.  This source appears to be extended enough that a significant portion of the total flux is lost on the largest angular scales by the filtering function of the interferometer.

While we attempted to fit the three data points with the most closely matched resolutions (our 3\arcsec~1 cm data at 23 mJy, the MAGPIS Epoch 1 5\arcsec~6 cm data at 19 mJy, and the 6\arcsec~20 cm data at 32 mJy) to a thermal bremsstrahlung spectrum, the uncertainties between the epochs and varying levels of recovered fluxes provide only a moderate constraint.  G$32.03+0.05$ has confirmed radio recombination lines from \citet{and11} at 91 \kms~consistent with being associated with the surrounding \nh3 gas.  The continuum source is also associated with 8 and 24 \micron~extended emission.  Additionally, in, e.g., Figure \ref{fig:hot_glm4_contours} we see morphological evidence that the continuum source has carved a bubble in the dense gas.  Despite the inconclusive thermal bremsstrahlung spectral fit, we conclude that this source is a young HII region of a massive star interacting with the surrounding dense gas.  From the relatively flat spectrum from 1 cm to 6 cm to 20 cm, we assume a turnover frequency (where the source becomes optically thick) at our longest observed wavelength, 20 cm.  A turnover at a longer (shorter) wavelength results in a lower (higher) emission measure and a later (earlier) spectral type.

Using the Str\"{o}mgren sphere approximation, we calculate the number of ionizing photons produced by this HII region and its main-sequence spectral type assuming that the turnover frequency occurs at 20 cm.  Our calculation of the number of ionizing photons per second produced by this HII region (given in the Appendix \S \ref{sec:strom}) gives Q = 10$^{47.8}$ s$^{-1}$, which corresponds to a spectral type B0.5 \citep{vac96}.

\section{The Commonality of Physical Conditions in Massive Star-Forming Regions}
\label{sec:litcomp}
The body of work observing pre- and star-forming clumps and characterizing their physical properties have led us to understand some common conditions.  
Together, existing high-resolution observations of massive star-forming regions show some common physical structure and properties of massive star-forming regions.  However, the extent of the innate similarities and differences between the observed regions are difficult to distinguish from differences in observing mode, resolution, or large-scale environment.  We place the current work in context by comparing with observations from the literature below and find some common features.  

Massive star-forming regions show clumpy filamentary structure on about 1 pc scales containing about 10$^{3}$ \Msun~of total mass \citep[e.g][]{wan08, pil11, fon12, liu12, gal10, gon10}.  Embedded within this filamentary structure are complexes of dense cores.  These cores are separated by about 0.1 pc \citep[e.g.,][]{rat08, pil11, fon12, dev11}, but higher resolution observations show further fragmentation, down to core separations of $\sim$0.01 pc \citep[e.g.,][]{gal10}.  The fragmentation scales observed are larger than the thermal Jeans Length \citep[e.g][]{pil11} implying the importance of turbulence in the fragmentation process.

Embedded within these complexes strung along pc-scale filaments are dense cores with masses of order 10-100 \Msun~\citep[e.g.,][]{pil11, liu12, bro09, zha09} and sizes of less than 0.007 pc at the highest resolution \citep[e.g][]{bro09} and around 0.1 pc at more moderate resolution \citep[e.g.,][]{dev11, wan07, rat08}.  Observations of the dense cores in massive star-forming regions give typical peak column densities of N(H$_{2}$) $\approx$ 5$\times$10$^{23}$ cm$^{-2}$ \citep[e.g.,][]{zha11, rat08, dev11, wan07} and up to 1$\times$10$^{24}$ cm$^{-2}$ at the highest resolution \citep{bro09}.  Many filaments show large-scale infall motions feeding filamentary gas toward dense clumps, as discussed in \citet{lon14}.

We find that pre-star-forming cores and filaments are characterized by smoothly varying temperatures of about 10-20 K \citep[e.g.,][]{zha09, zha11, rat08, pil11, rag11, dev11}.  These temperatures are also similar to those shown in the large study by \citet{san13}, who  also find a correlation between  the temperatures of cores and how clustered they are.  Interestingly, the pre-star-forming cores and filaments show very little structure in the temperature map and no temperature gradients toward the cores.  Star-forming cores uniformly contain warmer gas, about 40 K in this study and \citet{zha11}, but can be higher than 100 K when measured using higher excitation transitions of ammonia \citep[e.g.,][]{beu07, beu09}.
The massive star-forming region observed in this work shows two distinct kinematic components within the pc-scale filament.  Similar structures are also observed by \citet{dev11}, \citet{gal10}, and \citet{gon10}.  

While there are common masses and radii for cores identified in the literature, it is not clear at this point whether this corresponds to a common physical scale or simply a common observationally preferred scale.  While our \nh3 VLA observations have the potential to address this question over several clumps and core complexes, our observations still only cover one IRDC.  A systematic study of many regions would be required to understand the innate similarities and differences between the regions \citep[similar to the fragmentation analysis presented by][]{pal12}.

An interesting feature in this work found elsewhere in the literature is the trend that the coldest cores are more extended \citep[see Figure \ref{fig:radial} and e.g.,][]{zha11, li12, rat08, pil11}.  Since we model both the temperature and column density, this steepening in the column density profile with evolutionary stage is not a temperature effect.  Making the reasonable assumption that the coldest cores evolve over time to become more like the warmer cores suggests that massive star-forming cores begin larger and more extended, and gradually collapse and accrete material over time, quickly forming stars once the dense cores have formed.  This is consistent with the shallower intensity profiles observed in pre-star-forming clumps as seen in \citet{beu02}.  As the cores undergo internal collapse, their density profile steepens, and the dense cores quickly form stars.  An alternative explanation is that the difference in density profiles is due to the clumps eventual fate - a cluster containing massive stars starts with a steep profile, while a clustered low-mass SFR initially has a shallow density profile \citep{bon10}.

\section{The Large-Scale Environment of G32.02$+$0.06: A Massive Molecular Filament}
\label{sec:largescale}

\begin{figure*}
  \centering
  \includegraphics[width=1\textwidth]{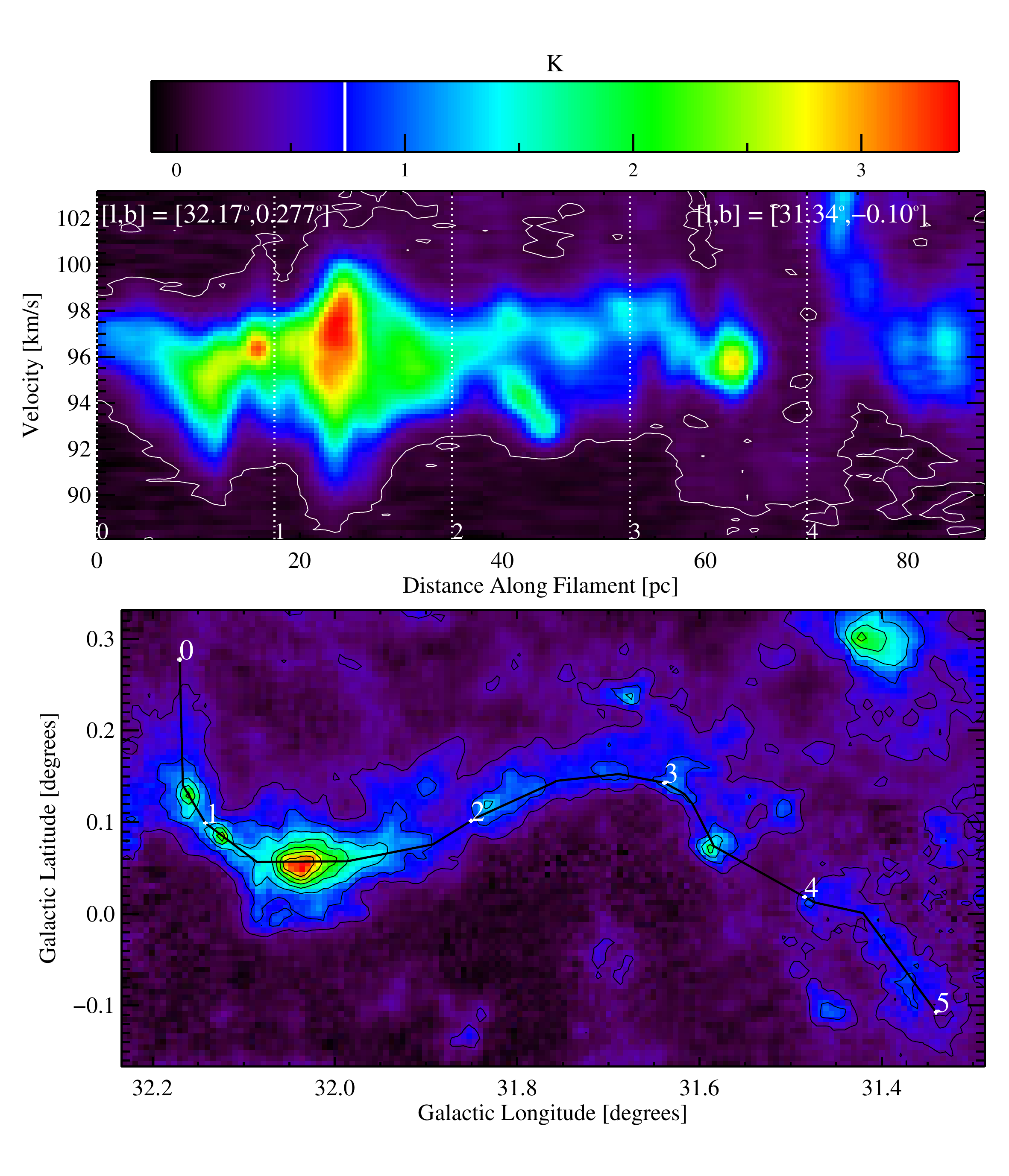}
  \caption{A position velocity diagram along the entire G32.09$+$00.09 GMC filament in the GRS \13CO cube (shown as the black line in the bottom panel superimposed on the \13CO integrated intensity from 88 to 102 \kms~map, numbers from 0 to 5 corresponding to dotted vertical lines shown along the x axis of the top panel).  This filament stretches over 60 pc with coherent velocity structure and contains a total mass of about 10$^{5}$ \Msun.
  }
    \label{fig:mmf}
\end{figure*}

\begin{figure*}
\centering
\includegraphics[trim = 8mm 5mm 7mm 5mm, clip, width=1\textwidth]{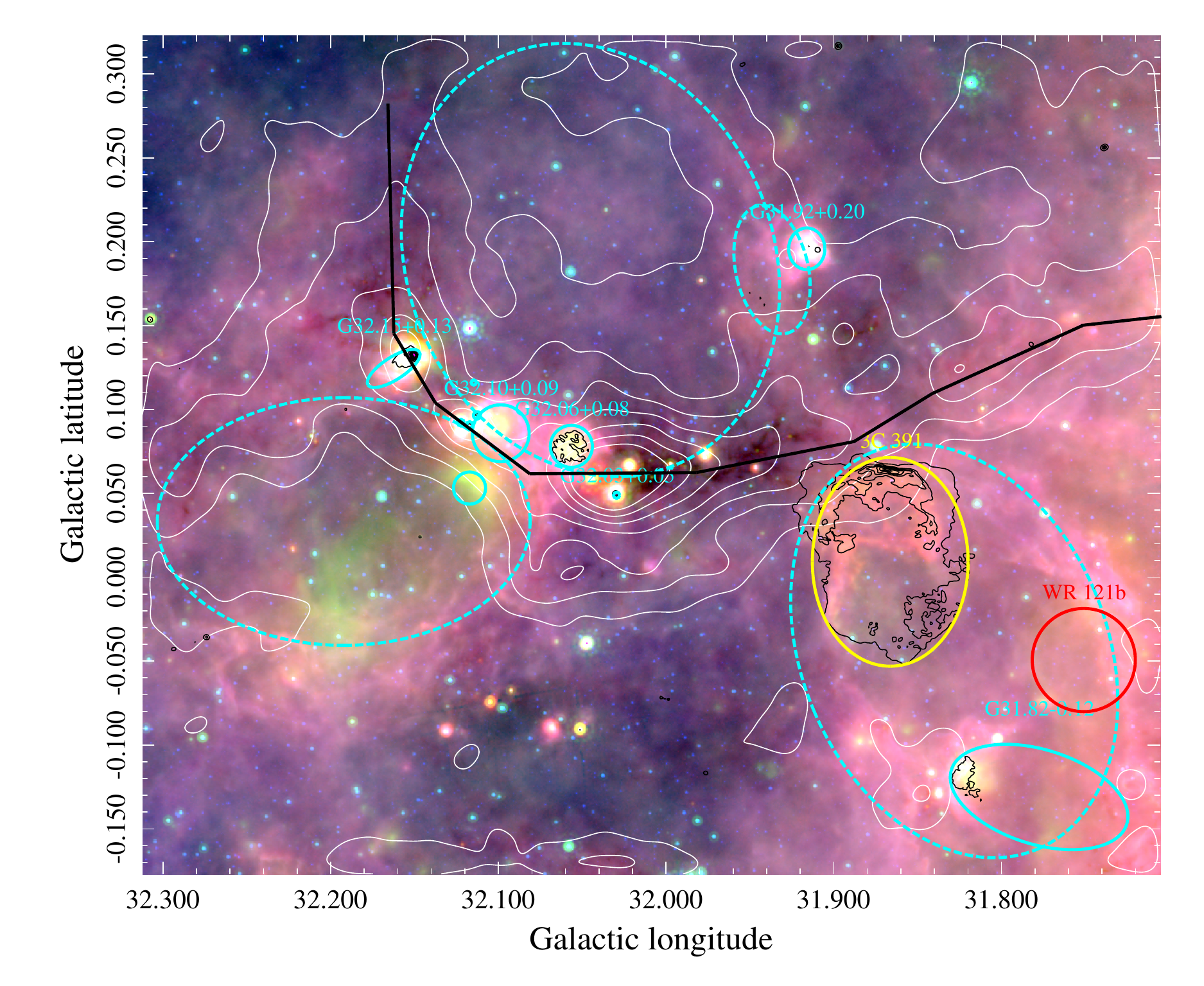}
\caption{The bubbles from previous generations of massive stars surrounding the G$32.02+0.06$ IRDC (image center).  We hypothesize that these bubbles shaped the filamentary molecular cloud ridge hosting the IRDC, seen as the \textit{white contours} (\13CO derived N(H$_{2}$) contours from 0.2 - 2 $\times$ 10$^{22}$ cm$^{-2}$).  The \textit{black contours} are 20cm continuum emission with levels from 2 - 100 mJy.  Bubbles and HII regions discussed throughout the text (see \S \ref{sec:largescale}) are shown as \textit{cyan circles} (dashed are bubbles, solid are HII regions), while the supernova remnant 3C 391 is highlighted with a \textit{yellow oval} and one of the shells of the Wolf-Rayet star WR 121b with a \textit{red oval}.  The three color image is composed of \textit{Red: 70 \micron, Green: 24 \micron, and Blue: 8 \micron.}  The \textit{black line} tracing the filament is the same as that shown in Figure \ref{fig:mmf}.}
\end{figure*}

The G$32.02+0.06$ filamentary IRDC is embedded within a much larger Giant Molecular Cloud (GMC), which is highly filamentary in nature.  Denoted as GMC 32.09$+$00.09 in the GRS cloud catalog, this cloud has a total mass of about 1.4 $\times$ 10$^{5}$ \Msun~\citep{rom10, rat09} and we classify it as a Massive Molecular Filament \citep[MMF;][]{bat12a} since it remains velocity coherent ($\Delta$v $<$ 5 \kms) over an 80 pc length and about 5 pc width (see Figure \ref{fig:mmf}).  The central velocity of about 95 \kms corresponds to a near kinematic distance of about 5.5 kpc.

Massive star formation has occurred in the immediate vicinity of G32.02$+$0.06.  This IRDC is surrounded by several HII regions, a young radio-bright supernova remnant, 3C391, and a Wolf-Rayet ring nebula surrounding WR 121b, spectral type WN7h \citep{gva10}. Below we investigate the interaction of these with the filament.

\subsection{HII regions inside and surrounding the GMC}

Compact HII regions include, in order of decreasing 20 cm surface
brightness and increasing size, G$32.03+0.05$, G$32.15+0.13$,  
G$32.06+0.08$, and G$32.10+0.09$ located along the warmest
dust at the northeast end of the G32.02$+$0.06 filament.  A cluster
of hyper-compact HII regions is located just north of G$32.10+0.09$.
Two of these HII regions, G$32.03+0.05$ (discussed in \S \ref{sec:freefree}) 
and G$32.06+0.08$, have confirmed radio recombination lines from \citet{and11, and12}.
These recombination lines are centered at about 91 and 96 \kms, respectively, clearly
associating the HII regions with the GMC complex.

Very dim and extended 20 cm continuum emission marks several much older 
bubbles.  
The bubble centered at  [$\ell$, \textit{b}] = $32.18+0.03$ has a major axis of about 13\arcmin~(21 pc) 
elongated along the Galactic plane.  This elliptical bubble is 
surrounded by warm dust emission at 8 and 24 \micron~on all sides.  
The bubble interior exhibits diffuse 24 \micron~emission typical
of dust in HII region.  Thus, this bubble is likely to mark
an old HII region whose western rim abuts GMC 32.09$+$00.09 
and compact HII region discussed above.  The northwestern
rim of this bubble is bright at 8 and 24 \micron, and can be seen
in CO around V$_{LSR}$ = 90 to 100 km~s$^{-1}$.
 
A second dim bubble centered at [$\ell$, \textit{b}] = $31.82-0.05$ is also rimmed
by an elliptical ring of warm 8 and 24 \micron~dust seen in emission.  
This bubble  has a major axis diameter of about 15\arcmin (24 pc) 
elongated orthogonal to the Galactic plane and encloses the SNR 3C391,
the WR 121b circumstellar shell, and the HII region G$31.82-0.12$.
As discussed below, there is strong evidence that the SNR is interacting
with the southern part of the G$32.09+00.09$ GMC and filamentary IRDC.

A smaller, 8 pc diameter bubble centered at
[$\ell$, \textit{b}] = $31.94+0.18$ is associated with a compact HII region
G$31.92+0.20$.  This compact HII region has radio recombination line
emission centered at about 100 \kms \citep{and11, and12} and is located
on the southern rim of
a CO shell at V$_{LSR}$ = 90 to 100 km~s$^{-1}$
and associated 8 and 24 \micron~ring seen in emission
centered at [$\ell$, \textit{b}] = $32.02, +0.21$ with a diameter of about 30 
pc. This ring is visible in the Hi-GAL dust continuum maps
between 70 to 500 \micron.  The northern rim of this ring
can be seen as an IRDC at 8 and 24 \micron.

Evidence that most of these structures are
physically related are provided by the bright 8 and 24 \micron~rims and IRDCs, associated
CO having radial velocities similar to the 
G32.02$+$0.06 filamentary infrared dark cloud, and 
closely matched radio recombination line emission.

\subsection{3C391 \& WR 121b}

The bright 3C391 SNR abuts the low-longitude side of the G32.02$+$0.06
IRDC \citep{bro05}.  This SNR is located in the interior
of the [$\ell$, \textit{b}]= $31.82-0.05$ bubble.  This bright remnant is interacting with
a molecular cloud at V$_{LSR} \approx$ 100 km~s$^{-1}$, similar
to the radial velocity of the G$32.09+00.09$ GMC \citep{rea96, rea98, rea99, wil98}.
Evidence for this interaction comes from the presence of OH
1720 MHz masers which trace the shock at the interface of the SNR
and the GMC \citep{fra96}.  

\citet{rea02} mapped the 2.12 \micron~and mid-IR 
lines of shock-excited H$_2$, 1.64 \micron~[FeII], and
several mid-IR lines such as 12 \micron~[Ne II] 
emission associated with 3C391.  The line emission
is brightest along the northwest rim of the SNR where it 
abuts the southwestern extension of the G$32.02+0.06$ IRDC.

WR 121b is located at a distance of 6.3$^{+2.2}_{-1.6}$ kpc, 
has a luminosity  $L \sim 10^{5.75}$ L$_{\odot}$, and a mass-loss 
rate  $\dot M \approx 10^{-4.7}$ M$_{\odot}$~yr$^{-1}$ 
\citep{gva10}.  The
Spitzer 24 \micron~images show a nearly perfectly round set of
four nested dust shells with radii of 30, 60, 100, and 140\arcsec
corresponding to radii of 0.9, 2, 3, and 4 pc.
The outer two shells are faintly visible in the Herschel
Hi-GAL 70 \micron.  Hints of these shells are also visible in the
Spitzer 8 \micron~MIPS images.

\citet{gva10} proposed that WR 121b might be a massive
run-away star ejected from from the HII region W43, located about 1\deg~(110 pc) from this star.  
They hypothesize that the massive binary star WR 121a consisting of
a $WN7+OB$ star binary in the central cluster ionizing 
W43 and the star ALS 9956 may have been associated with the
disintegration of a non-hierarchical systems of massive stars.  Assuming
that the star has traveled $\approx$ 110 pc in less than 5 Myr (the 
main-sequence lifetime of a progenitor to a massive WR-star), implies
a velocity $V \ge 20$ \kms.  If WR 121b is a run-away star, 
it would be expected to produce a bow-shaped nebula as its massive 
star winds interact with the  surrounding ISM.  

The nearly completely circular morphology of the circumstellar nebula
surrounding WR 121b strongly suggests that it is nearly stationary
with respect to the surrounding ISM.  It is possible that both 3C391 and
WR 121b and its circumstellar shells are embedded within the 
 [$\ell$, \textit{b}] = $31.82-0.05$ bubble.

The G32.02$+$0.06 GMC is a filamentary structure sandwiched between several
old, 10 to 50 parsec diameter bubbles created by older generations
of massive stars.  Assuming that these HII regions are expanding a
typical speeds of order 5 \kms~implies that their ages
are between 2 to 10 Myr.  The string of compact and hyper-compact
HII regions between G$32.03+0.05$ and G$32.10+0.09$ occupy the wormiest, condensed and twisting,
part of this GMC. This region may have been compressed on both 
sides by the two bubbles centered at [$\ell$, \textit{b}] = $32.19-0.03$ and
[$\ell$, \textit{b}] = $32.02, +0.21$.  

\subsection{Interpretation}

The host GMC to the IRDC G32.02$+$0.06 appears to be a nearly 100 pc
long by $\sim$5 pc wide filamentary cloud which has been shaped by 
older generations of massive stars.  At least three 10 to 30 pc 
diameter bubbles, likely old HII regions, each of which contain at 
least several massive stars appear to have compressed this cloud
and created its various loops and bends.  The region of greatest
star formation activity between l = 32.0 and 32.2 seems to have
been compressed on both sides by a pair of bubbles.  The older bubble is
located due west of this part of the filament while the young bubble
is located due east.  

The darkest portion of this IRDC is located directly south of 
this active ridge.  The southern end of this cloud appears to be
interacting directly with the bubble which hosts the SNR 3C391
and the Wolf-Rayet star WR 121b.  The presence of these two objects
likely indicates the presence of an older OB association in this
region.

\section{Conclusions}
\label{sec:conclusion}
We probe the physical conditions at the onset of massive star formation and how these properties evolve.  By observing a range of evolutionary states toward a single IRDC (G32.02$+$0.06), we are able to make a direct comparison of the physical properties of the evolutionary states, without the usual observational biases introduced by differing distances/resolutions, Galactic-scale abundance variations, and large-scale environment.

We observed two regions of the IRDC G32.02$+$0.06 with the (1,1), (2,2), and (4,4) inversion transitions of para-\nh3 on the VLA ($\sim$0.1 pc resolution) and derived temperatures and column densities, velocities, and velocity dispersions using a radiative transfer model.  The observed properties are compared with the literature to explore the range of physical conditions observed in massive star-forming regions.  We also investigate the large-scale environment of the Massive Molecular Filament hosting this IRDC and suggest that it has been shaped by previous generations of massive stars.

\begin{itemize}
\item \textit{Temperature Structure:} The temperature structure in the pre-star-forming cores and filaments is smooth, varying from about 10 to 20 K.  The filaments show no gradient toward the cold cores, except Core 7 in the active clump which shows a temperature gradient from 20 K to 15 K.  There is a slight gradient from high to low temperature from the filament exterior to its interior.  The pockets of active star formation (the warm core complex, especially Core 2 in the active clump) show an increase in the temperature, to about 35-40 K.  

\item \textit{Column Density Structure and the Evolution of Massive Star-Forming Clumps:} Both clumps show core sub-structuring, but different density profiles.  The quiescent clump contains a smooth, extended central core at the hub of the filaments.  This core is over 0.2 pc in diameter, cold, with a peak surface density of about 0.5 g cm$^{-2}$.  This core shows no signs of active star formation \citep[Stage 1;][]{bat10} and may represent an early stage in massive star formation.  The active clump shows three complexes of compact, dense cores near the fiducial \citet{kru08} 1 g cm$^{-2}$ threshold for massive star formation.  These three complexes show very similar density structures (similar core sizes, separations, and peak surface densities).  However, the warm core complex is in Stage 3 \citep{bat10} and appears to be actively forming massive stars, while the dark core complexes are in Stage 2 \citep{bat10}, showing the signs of early, embedded massive star formation (8 \micron~dark and cold, with 6.7 GHz CH$_{3}$OH maser emission).  Abutting the dark core complex to the south is a young UCHII region \citep[Stage 4;][]{bat10} which has carved a bubble in the \nh3 gas.

Given the similarity of the column density structures observed, we hypothesize that we are seeing core complexes in slightly different evolutionary stages, from the dark core complexes (Stage 2) to the warm core complex (Stage 3) to the young UCHII Region (Stage 4).  Whether or not the Stage 1 extended cold core observed in the quiescent clump represents the precursor to the dark core complexes remains an open question.  There is  support in the literature to suggest that cold cores are generally more extended and less massive, implicating the process of global collapse and accretion in the formation process of dark core complexes.  A conclusive answer would require a larger-scale systematic study.

\item \textit{Virial Parameters and Kinematics:} In our sample of 16 identified cores, the virial parameters are low (average of 0.6, max. of 1.6), suggesting that turbulence and thermal pressure support are insufficient to support the cores against gravitational collapse.  Core masses range from about 10-100 \Msun~and average line widths are about 0.5 -1 \kms.  We present position velocity diagrams along the star-forming filaments and find that they show a smooth velocity structure punctuated by discontinuities at the sites of active star formation.  The large core in the quiescent clump may be at the interaction layer between two kinematically distinct filaments.  In the eastern edge of the quiescent clump, these kinematically distinct filaments masquerade as a single filament in projection.  This superposition is a concern for anyone studying filaments through their continuum emission alone, as noted by \citet{moe14}.

\item \textit{Common Conditions for Massive Star Formation:}  We compare our results with previous work in the literature and uncover some common conditions for the formation of massive stars.  We find clumpy filamentary structure on $\sim$1 pc scales with massive cores, extended ($>$0.1 pc) in the earliest phase and compact ($<$0.1 pc) in later stages.  Commonly derived fragmentation scales are larger than the thermal Jeans lengths, indicating that turbulence is important in the fragmentation process in these cores.  The onset of massive star formation disrupts filamentary structure.  The pre-star-forming filament and cores show little temperature variation and average about 10-20 K, while the star-forming cores have a higher average temperature, with a peak near 40 K.

\item \textit{The Massive Molecular Filament hosting the G32 IRDC:}  The G32 IRDC is embedded within a much larger MMF, which shows velocity coherence ($\Delta$v $<$ 5 \kms) over 60 pc in length.  The total mass of the MMF is about 10$^{5}$ \Msun~and its width is about 5 pc.  The velocity coherence along the filament may indicate that it is nearly in the plane of the sky, or if not, any flows along the filament are small/non-accelerating.  There is evidence for bubbles above and below the MMF from previous generations of massive stars.  These bubbles may have shaped the twists and bends observed along the filament's length, as well as possibly having a role in the compression of a GMC into a dense, molecular filament.

\end{itemize}

The observations presented allow us to investigate the role of evolution independently of Galactic-scale abundance variations, distance/resolution and large-scale environmental effects toward one IRDC showing different evolutionary stages.  We identify common conditions observed toward massive star-forming regions and discuss how these physical structures may evolve with time.  In a companion paper \citep{bat14a}, we derive the abundance, compare gas and dust properties, and discuss the origin of millimeter continuum emission.

\acknowledgments

We thank the referee for his or her comments, which have improved the quality of this manuscript.  We thank Rick White and the MAGPIS team for their assistance understanding continuum fluxes of the free-free source.  We thank Roberto Galv\'an Madrid for comments on an early draft of this paper.
This work has made use of the GLIMPSE and MIPSGAL surveys, and we thank those teams for their help and support.  We would like to thank the staff at VLA for their assistance.  The National Radio Astronomy Observatory is a facility of the National Science Foundation operated under cooperative agreement by Associated Universities, Inc.  This publication makes use of molecular line data from the Boston University-FCRAO Galactic Ring Survey (GRS). The GRS is a joint project of Boston University and Five College Radio Astronomy Observatory, funded by the National Science Foundation under grants AST-9800334, AST-0098562, AST-0100793, AST-0228993, \& AST-0507657.  This work has made use of ds9 and the Goddard Space Flight Center's IDL Astronomy Library.  Data processing and map production of the Herschel data has been possible thanks to generous support from the Italian Space Agency via contract I/038/080/0. Data presented in this paper were also analyzed using The Herschel interactive processing environment (HIPE), a joint development by the Herschel Science Ground Segment Consortium, consisting of ESA, the NASA Herschel Science Center, and the HIFI, PACS, and SPIRE consortia. 

\begin{appendices}
\section{HII Region Spectral Type Identification}
\label{sec:strom}
We approximate the number of ionizing photons produced in the HII region G32.03$+$0.05 using the Str\"{o}mgren sphere approximation.  The Str\"{o}mgren sphere  approximation assumes that the HII region is spherical, in equilibrium, that there is one electron per ion, and that each energetic photon ionizes one atom.  We calculate the radius of the HII region using the expression
\begin{equation}
{\rm r} = \bigg[\frac{{\rm S}_{\nu}{\rm c}^{2}}{2\nu^{2}{\rm k}{\rm T}_{{\rm e}}}4{\rm D}^{2}\bigg]^{1/2}
\end{equation}
from which we derive a radius of 0.08 pc at 20 cm.  This matches our empirical Gaussian fit to the source, which gives a source radius of 0.1 pc.  In the expression above, S$_{\nu}$ is the flux density, $\nu$ is the observed frequency, T$_{e}$ is the electron temperature which we take to be 8000 K, and D is the distance which we assume to be 5.5 kpc.  Assuming that the source becomes optically thick at our longest observed wavelength where the spectrum appears relatively flat, $\lambda$ = 20 cm, we can derive an emission measure using the expression \citep[from, e.g.,][]{woo89}
\begin{equation}
{\rm EM} [ {\rm pc}~{\rm cm}^{-6}] = \frac{\tau_{\nu}}{8.235\times10^{-2}\bigg(\frac{{\rm T}_{{\rm e}}}{{\rm K}}\bigg)^{-1.35}\bigg(\frac{\nu}{{\rm GHz}}\bigg)^{-2.1}a(\nu,{\rm T}_{{\rm e}})}
\end{equation}
where $\nu$ is the turnover frequency (where $\tau$=1) which we take to be 1.5 GHz (20 cm) and a($\nu$, T$_{e}$) is a correction factor of order unity.  We can then calculate the number density of electrons using the expression
\begin{equation}
{\rm n}_{{\rm e}} = \sqrt{\frac{{\rm EM}}{2{\rm r}}}
\end{equation}
and assume that this equals the number density of ions.  The number of Lyman continuum photons is given by the Str\"{o}mgren sphere equation,
\begin{equation}
{\rm Q} = \frac{4}{3} \pi {\rm r}^{3} \alpha_{B} {\rm n}_{e}^{2}
\end{equation}
where $\alpha_{B}$ is the case B recombination rate coefficient, 3.1 $\times$ 10$^{-13}$ cm$^{3}$ s$^{-1}$ for T$_{e}$ = 8000 K.  We derive 6.4 $\times$ 10$^{47}$ Lyman continuum ionizing photons per second, which corresponds to a spectral type of B0.5 \citep{vac96}.

\end{appendices}

\bibliography{/Users/battersby/Dropbox/references1}{}

\end{document}